# Dynamics of Current, Charge and Mass

**Bob Eisenberg**

Department of Applied Mathematics
Illinois Institute of Technology
USA

Department of Physiology and Biophysics
Rush University
USA
bob.eisenberg@gmail.com

**Xavier Oriols**

Departament d'Enginyeria Electrònica, Universitat Autònoma de Barcelona,
SPAIN
xavier.oriols@uab.es

**David Ferry**

School of Electrical, Computer, and Energy Engineering
Arizona State University
USA
ferry@asu.edu

*September 8, 2017*

## ABSTRACT

Electricity plays a special role in our lives and life. The dynamics of electrons allow light to flow through a vacuum. The equations of electron dynamics are nearly exact and apply from nuclear particles to stars. These Maxwell equations include a special term, the displacement current (of a vacuum). The displacement current allows electrical signals to propagate through space. Displacement current guarantees that current is exactly conserved from inside atoms to between stars, as long as current is defined as the entire source of the curl of the magnetic field, as Maxwell did. We show that the Bohm formulation of quantum mechanics allows the easy definition of current without the mysteries of the theory of quantum measurements. We show how conservation of current can be derived without mention of the polarization or dielectric properties of matter. We point out that displacement current is handled correctly in electrical engineering by 'stray capacitances', although it is rarely discussed explicitly.

Matter does not behave as physicists of the 1800's thought it did. They could only measure on a time scale of seconds and tried to explain dielectric properties and polarization with a single dielectric constant, a real positive number independent of everything. Matter and thus charge moves in enormously complicated ways that cannot be described by a single dielectric constant, when studied on time scales important today for electronic technology and molecular biology. When classical theories could not explain complex charge movements, constants in equations were allowed to vary in solutions of those equations, in a way not justified by mathematics, with predictable consequences.

Life occurs in ionic solutions where charge is moved by forces not mentioned or described in the Maxwell equations, like convection and diffusion. These movements and forces produce crucial currents that cannot be described as classical conduction or classical polarization. Derivations of conservation of current involve oversimplified treatments of dielectrics and polarization in nearly every textbook. Because real dielectrics do not behave in that simple way—not even approximately—classical derivations of conservation of current are often distrusted or even ignored. We show that current is conserved inside atoms. We show that current is conserved exactly in any material no matter how complex are the properties of dielectric, polarization, or conduction currents.

Electricity has a special role because conservation of current is a universal law. Most models of chemical reactions do not conserve current and need to be changed to do so. On the macroscopic scale of life, conservation of current necessarily links far spread boundaries to each other, correlating inputs and outputs, and thereby creating devices. We suspect that correlations created by displacement current link all scales and allow atoms to control the machines and organisms of life. Conservation of current has a special role in our lives and life, as well as in physics.

We believe models, simulations, and computations should conserve current on all scales, as accurately as possible, because physics conserves current that way. We believe models will be much more successful if they conserve current at every level of resolution, the way physics does. We surely need successful models as we try to control macroscopic functions by atomic interventions, in technology, life, and medicine.

Maxwell's displacement current lets us see stars. We hope it will help us see how atoms control life.

# 1. INTRODUCTION

The dynamics of electrons allow us to hold a computer in our hand that detects signals of microvolts, from a 500 watt satellite source some 22,200 miles away. The computer in our hand makes logical decisions nearly a billion times a second, using some $10^{12}$ components, with hardly any errors.

The fundamental laws that govern these phenomena are Maxwell's equations. These laws are so general that they are often thought to have limited practical applicability: their application is often thought to depend on precise knowledge of the detailed properties of matter, knowledge that is often unknown, always hard to acquire. This paper is about a notable exception: conservation of current. ***Conservation of current is true universally, on all scales, independent of the detailed properties of matter.***

Kirchhoff's current law illustrates the importance of conservation of current. Kirchoff's laws use a set of currents and voltages to predict the performance of systems operating with currents ranging from femtoamps to kiloamps, with potentials from microvolts to hundreds of volts, in resistors ranging from tenths of an ohm to sometimes tens of gigohms. Kirchoff's laws are simple, compact and easy to use. They are also exact in branched one dimensional systems, when current is defined to include displacement current. Electrical systems follow Kirchoff's current law exactly because conservation of current is universal.

***Electricity is Different*** because few physical systems follow simple and compact laws with such precision.

Electricity is familiar as well as different because it is used so widely in our technology and life. Our society of information (with its internet of everything) is a practical application of the dynamics of electrons. Our technology would be impossible if Kirchoff's laws were not accurate and easy to apply. Electricity is so widely used because it follows universal laws that can be easily applied.

Compact and simple laws, like Kirchhoff's laws, allow the use of mathematics to design devices with a wide range of properties (Gray, Hurst et al. 2001, Cressler 2005, Horowitz and Hill 2015) with reasonable realism. For example, the microchip in your laptop computer requires manufacturing precision to sub-nanometer accuracy across 300 millimeters of the semiconductor wafer in which the computer chip is formed. This accuracy is an incredible feat of today's technology.

Sciences that depend on less accurate, simple and compact laws are often forced to use models that are not 'transferrable' (as the word is used by chemists). We mean by 'transferable' that the same law—with the same numerical value of parameters—can be used in a multitude of conditions and systems and is not constrained to a single system and set of conditions. Non-transferrable models use parameters that change with conditions, often in ways that are hard to capture or predict. Devices become difficult to use when their parameters and properties vary in unpredictable ways.

Nearly all systems — particularly liquids and ionic solutions so important in chemistry and biology — involve many types of forces and interactions. Interacting systems are particularly difficult to capture in simple and compact laws. Interactions make it difficult to find transferrable models, with one set of unchanging parameters valid for a large range of conditions. The simple and compact transferable models valid for typical electronic technology cannot be automatically applied to biological systems because of their complex structure, but the electrical properties of individual nerve and muscle fibers can be expressed in terms of Kirchoff's laws and little else, amazingly enough. (Hodgkin and Huxley 1952c, Hodgkin 1958, Hodgkin 1964, Hodgkin 1992, Weiss 1996, Huxley 2000, Huxley 2002, Prosser, Curtis et al. 2009, Gabbiani and Cox 2010). Even electrical syncytia like the heart, epithelia, lens of the eye, liver, and so on can be described quite well with modest generalizations of Kirchoff's laws.(Tung 1978, Eisenberg, Barcilon et al. 1979, Mathias, Rae et al. 1979, Eisenberg and Mathias 1980, Mathias, Rae et al. 1981, Geselowitz and Miller 1983, Levis, Mathias et al. 1983)

Nerve and muscle fibers live in salt solutions derived from seawater, as does nearly all of life. Many chemical systems and a great deal of our chemical technology involve these salt solutions. Interactions abound in salt solutions, and they occur between the different types of ions, and ions with the water. Seawater flows in pressure and temperature gradients, so many types of forces are involved. Electric fields are particularly important in these systems and they pose particular problems because electric fields are very strong and extend a very long way, coupling atomic and macroscopic length scales with one set of physical laws.

Viewed physically, most biological systems of interest are macroscopic systems containing a huge number of fundamental particles with a fantastic number of interactions between pairs of particles. The number of interactions is orders of magnitude larger than Avogadro's number or $10^{23}$ for the number of particles per mole. Even small systems contain millions of molecules, and larger systems contain $10^{17}$ molecules, pairwise interactions can dominate properties. The attempts to describe the system by computing the dynamics of each particle becomes, in general, computationally impossible when these number of interactions are involved.

Some general properties about the behavior of biological systems are controlled by a handful of atoms, as molecular biology has so well shown us, and the role of those atoms must be understood at such an atomic level.(Alberts, Bray et al. 1994) But that understanding does not require computation of all atoms or all interactions. In some tissues, like nerve and muscle cells, and some syncytia, already described, electrical properties of cells and tissues on the macroscopic scale are understood nearly completely from atomic properties and structures. The link between atoms and cells is known and turns out to be a slight generalization of the same Kirchoff's laws that are so important in the design of our technology.

In this paper, we show that electrical current satisfies a current conservation law exactly and universally when it includes an additional component beyond the flux of charge: the displacement current (Zapolsky 1987, Arthur 2008, Selvan 2009, Arthur 2013). The displacement current plays a crucial role in the practical application of Kirchoff's laws.

The fact that the modeling of systems with charged particles has to include both particle current and displacement current, rather than only particle current, is a main message of this paper.

At first sight, the message may seem trivial. It is clearly explained in most elementary textbooks. However, there is a surprisingly large amount of relevant work presently being published in biology, electronics, chemistry, etc., where the dynamics of charged systems are described but the displacement current generated by the movement of charge is ignored. Indeed, it was a surprise to find important work which ignores current flow altogether.(Eisenberg 2014a,b) It seems to us that emphasizing the importance of displacement current is still necessary in the scientific community. And we hope that including displacement current will make models more useful, transferable, and realistic.

## 1.1 The strength of the electric field in life

Electric forces are much stronger than other forces we deal with in ordinary life (e.g., in mechanical systems, diffusion in liquids, and heat flow). One per cent changes in concentration, or mass density or temperature have little effects in ordinary life. One per cent errors in the computation of heat flow, convection, or diffusion are not very good, but are not a disaster either. But a one percent change in the source of the electric field has dramatic effects: as Feynman says in the third paragraph of his textbook on electrodynamics (Feynman, Leighton et al. 1963), one per cent of the charge in a person (at a distance of 1 meter) creates a force large enough to lift the earth. Indeed, such forces are large enough to ionize the atoms around and in us, ionizing them into a gaseous plasma, destroying us and our laboratories in a significant explosion. In normal life, most people have seen sparks at electrical outlets and have seen and heard lightning. It takes only an easy calculation to learn that there is a tremendous amount of energy being dissipated from the clouds during the storm. Life and biological experiments are compatible with only tiny changes in charge density, closer to $10^{-15}$ than a $10^{-2}$ fraction of all charges present. For example, a modern microcomputer in your cell phone involves transistors that switch with only about a thousand electrons ($\sim 10^{-16}$ Coulombs), a vanishing fraction of the total number of electrons in the transistor.

Electric forces are so strong that they change the shape of things, much as the gravitational force of the moon distorts the shape of the earth by moving our oceans and creating tides. Similarly, electric forces change the distribution of charge, in a way called polarization. Indeed, early workers in electricity (Faraday and Maxwell) and JJ Thomson, (before he discovered the electron, see (Thomson 1893)) were aware of polarization and only dimly aware of charge. A search of Thomson (Thomson 1893) does not find the word charge anywhere in the book. Evidently, Thomson did not know of permanent charge independent of the electric field (Buchwald 1985) until he discovered the electron (Thomson 1898, Thomson 1906).

## 1.2 The current conservation law in electrical circuits

Computers as we know them are possible because Kirchoff's laws of electrical networks are robust subsets of the universal laws of electrodynamics that accurately describe the properties of circuits. Our computers are built almost entirely of circuits in which current flows in one dimension in wires and devices (like resistors, capacitors and field effect transistors). Circuits are almost always branched networks of one-dimensional components. Currents at branch points ('nodes') add and subtract so total current is conserved exactly, always, at all times. Everything coming into a node goes out of the node, as described by Kirchoff's current law. In Table 1, we have defined the four types of current discussed in this paper. The magnitude $\mathbf{J}_{mass}$ refers to the flux of mass. $\mathbf{J}_Q$ is the flux of charge. In Table 1, we also include the new displacement current:

$$\mathbf{J}_D = \varepsilon_0 \, \partial\mathbf{E}/\partial t \,, \tag{1}$$

where **E** is the electric field, $\varepsilon_0$ is a constant, the permittivity of a vacuum, that never changes with anything, and $t$ is time. Finally, the total current $\mathbf{J}_{total}$ is defined as the sum of the charge and displacement current:

$$\mathbf{J}_{total} = \mathbf{J}_D + \mathbf{J}_Q. \tag{1.b}$$

The total current $\mathbf{J}_{total}$ that enters a node, leaves it. Total current[1] $\mathbf{J}_{total}$ is ***exactly*** equal ***everywhere at every time in every device*** in a series circuit, even though the charge transport (the flux) $\mathbf{J}_Q$, can be very different in each device, as different as charge transport in a wire is from that in seawater, or from the displacement current in a 'vacuum' capacitor $C\frac{\partial V}{\partial t}$ (coulombs per sec, SI official name $Cs^{-1}$) where $V$ is potential in volts (SI official name *V*), $t$ is time. The capacitance $C$ is in farads.[2]

**Table 1: Flux**

| Name | Nickname | Symbol | units SI |
|---|---|---|---|
| Flux of Mass | Flux | $\mathbf{J}_{mass}$ | kg $s^{-1}$ $m^{-2}$ |
| Flux of Charge | Current of charge *or (sadly)* Current | $\mathbf{J}_Q$ | C $s^{-1}$ $m^{-2}$ |
| Displacement Current | $\varepsilon_0 \, \partial E/\partial t$ | $\mathbf{J}_D$ | C $s^{-1}$ $m^{-2}$ |
| Total Current | Total Current | $\mathbf{J}_{total}$ | C $s^{-1}$ $m^{-2}$ |

Consider a circuit with a battery connected in series, through a wire, to a resistor and a capacitor. Although the physics of charge movement is entirely different in a battery, wire, resistor, or vacuum capacitor, the total current is exactly equal at all times in all positions of the series circuit and under all conditions. Eisenberg (2016c: Fig. 2) describes this reality in some detail.

---

[1] We assume that the cross-sectional area is constant in this paragraph so that we do not have to distinguish between current ***I*** and current density **J** (or current per unit area).

[2] Note that if the potential is a sinusoid, say $V(t) = \sin \omega t$, as it is in the enormous classical literature measuring polarization currents and dielectric 'constants', the current through the capacitor is $C\frac{\partial \sin \omega t}{\partial t} = C \, \sin(\omega t - 90^\circ)$. The current through a perfect capacitor is 'ahead' of voltage by a phase angle of $90^\circ$.

The total current is $\mathbf{J}_{total}$ and it is hard to accept that this will be exactly conserved when so many mechanisms are involved over such a range of times and forces. Yet it is. How is it possible for current $\mathbf{J}_{total}$ to be exactly conserved in a series circuit, independent of the mechanisms of charge transport $\mathbf{J}_Q$, from say $10^{-16}$ sec to $10^2$ sec, and from $10^{-6}$ volts to $10^2$ volts (and very much larger)? This conservation is just a consequence of Maxwell's equations as will be demonstrated in section 2.2. It can also be understood as a consequence of a particle conservation law for particles if the particles have charge and therefore satisfy Gauss' law. Without electricity and Gauss' law, particle flux $\mathbf{J}_{mass}$ would be conserved in a series hydraulic circuit of (say) water pipes. With electricity and Gauss' law, particle flux $\mathbf{J}_Q$ is NOT conserved in a series circuit of say resistors. Current $\mathbf{J}_{total}$ is conserved but not particle flux $\mathbf{J}_Q$. Currents are exactly equal in a series circuit because total current $\mathbf{J}_{total}$ has another component beyond the flux of charge $\mathbf{J}_Q$ (coulombs per second) associated with the flux of mass $\mathbf{J}_{mass}$ (units kilograms per second per $m^2$). The other component of the conserved total current $\mathbf{J}_{total}$ is Maxwell's displacement current $\mathbf{J}_D = \varepsilon_0\, \partial\mathbf{E}/\partial t$ of Eq. (1). The displacement current $\mathbf{J}_D$ depends only on $\partial\mathbf{E}/\partial t$. It does not depend on the properties of matter or its dielectric coefficient $\varepsilon_r$ (dimensionless) because we use $\varepsilon_0$ in the definition of displacement current. The displacement current we define does not depend on properties of matter. $\mathbf{J}_D$ is different from $\mathbf{J}_{total}$ and from $\mathbf{J}_{mass}$. Displacement current is determined only by the rate of change of the electric field and not by any property of matter whatsoever. $\mathbf{J}_D$ is not produced by the mechanisms that determine $\mathbf{J}_Q$ and $\mathbf{J}_{mass}$. Indeed, it must be clearly understood that the flux of charge or mass inside a capacitor is zero.

$$\overbrace{\mathbf{J}_Q = 0;\ \mathbf{J}_{mass} = 0;\quad \mathbf{J}_{total} = \mathbf{J}_{\boldsymbol{D}}}^{\textit{Inside a capacitor}}$$

(Zapolsky 1987, Arthur 2008, Selvan 2009, Arthur 2013) have particularly useful discussions of displacement current $\mathbf{J}_D$, and we will discuss it in great detail below.

We see then that the **electric field changes to ensure perfect equality of total current everywhere in everything at every time in a series circuit,** as a solution of Maxwell's equations of electrodynamics. Biological systems are usually modelled in a three dimensional physical space. The one dimensional model is applicable to the nerve and muscle cells already discussed and easily generalized to syncytia like the heart. In any case, we will see in section 2 that the conclusions mentioned above about the importance of the total current (with particle and displacement components) can be directly extrapolated to three dimensional systems in general.

The charge density carried by mass density can be a complex function reflecting the multifaceted distribution of charge in matter on all scales and so is described by many parameters and variables, all of which can interact with each other. A model and theory of matter and its charge is needed to relate mass and charge density. The theory must include dynamics to derive the movement of charge $\mathbf{J}_Q$ from the movement of mass $\mathbf{J}_{mass}$. Many components may be involved, of different chemical species, concentration, and molecular/atomic charge per chemical species (i.e., 'valence' of atomic or molecular ion). The dynamics of each component may depend on many types of forces and fields, electrical and convection to be sure, but also diffusional, thermal, and gravitational for example. Most importantly, the dynamics of one component is usually coupled to the dynamics of another. If the components are charged, they are coupled by the electric field. If the components have finite size, they are coupled by steric forces because a certain number of finite size components fill space. Components interact so they cannot overfill space. Interactions are not local; indeed, electrical interactions always involve spatial boundary conditions because they are described by partial differential equations, field theories that in general extend to infinity sometimes with unexpected results (Mertens and Weeks 2016). Steric forces are not local, although they need not reach infinity or extend to far boundaries. In general 'everything interacts with everything else' in many ways and by many interactions specific to each system of interest.

In spite of the fact that the four Maxwell's equations, together with the dynamical laws of movement, can be compactly written in a small piece of paper, it is obviously impossible to solve them all to have a general model and theory of matter and its charge. We shall see however that the fundamental principles of conservation of total current $\mathbf{J}_{total}$ and charge $Q$ can be applied to all matter, no matter what the relation of the movement of charge $\mathbf{J}_{\mathrm{Q}}$ and the movement of mass $\mathbf{J}_{mass}$ . Application of these principles leads to practical results important in the understanding and design of engineering and biological systems.

## 1.3 Polarization Charge and Current

The charge density $\rho_{\mathrm{Q}}$ carried by mass density[3] $\rho$ can be a complex function reflecting the multifaceted distribution of charge in matter on all scales (from nuclear to atomic to molecular to macroscopic, including interface conditions and boundary conditions) and so is described by many parameters and variables, all of which can interact with each other. A model and theory of matter and its charge is needed to derive $\rho_{\mathrm{Q}}$ from $\rho$. The theory must include dynamics to derive the movement of charge $\mathbf{J}_{\mathrm{Q}}$ from the movement of mass $\mathbf{J}_{mass}$. Many components may be involved, of different chemical species, concentration, and molecular/atomic charge per chemical species (i.e., the charge number of atomic or molecular ions nicknamed 'valence' in classical chemistry). And the dynamics of each component may depend on many types of forces and fields, electrical and convection to be sure, but also diffusional, thermal, and gravitational for example. Most importantly, the dynamics of one component is usually coupled to the dynamics of another. If the components are charged they are coupled by the electric field. If the components have finite size, they are coupled by steric forces because a certain number of finite size components fill space.

Of course, some of that movement of mass and its charge in a resistor is much more complicated. In an atom, for example (or for a molecule), the bound electrons can move differently from the nucleus. The electrons carry negative charge while the nuclei carry positive charge. This can result in a displacement between the positive and negative charge, either permanently or in response to the electric field. The displacement will be very different at different times and locations. This kind of movement is conventionally called polarization or more exactly polarization current. Polarization current can be called dielectric displacement current if it behaves 'well' and follows the physical law $(\varepsilon_r - 1)\varepsilon_0\, \partial \mathbf{E}/\partial t$ with $\varepsilon_r$ being a real positive constant called the dielectric constant (> 1), independent of time and **E.** Such idealized dielectric ***constants*** and polarization currents exist in textbooks and models. *They do not exist in matter* and assuming that matter behaves in this naïve (and unrealistic) way can lead to serious errors and misunderstandings.

[3] It is unfortunate that the same symbol is normally used for two different quantities—mass density and charge density, but we shall try to be specific at the various points where confusion may arise. They must both appear as it is possible that some of the mass is, in fact, charge neutral and will not appear in the equations for charge.

Polarization currents have a large and striking dependence on time in almost all materials, even in the solid phase, and is a main subject of classical work (Debye and Falkenhagen 1928, Fuoss 1949, Fröhlich 1958, Van Beek 1967, Nee and Zwanzig 1970, Böttcher, van Belle et al. 1978, Barthel, Buchner et al. 1995, Kurnikova, Waldeck et al. 1996, Buchner and Barthel 2001, Heinz, van Gunsteren et al. 2001, Kremer and Schönhals 2003, Rotenberg, Dufre Che et al. 2005, Kuehn, Marohn et al. 2006, Angulo-Sherman and Mercado-Uribe 2011). The practical importance of the time dependence is well known to the engineers who design solid state devices that work. Ch. 6 of (Hall and Heck 2011) gives a clear description of polarization in real materials, showing that the classical approximation of a dielectric constant (as a single real number) is of little use. Their analysis of a harmonic element of a classical harmonic oscillator—a charged mass on a spring with dashpot (Fig 6-5 p. 258)—is particularly revealing. No one would approximate the location of a mass on a spring as a time independent constant if they could avoid it. Obviously, the mass and its charge will move in most situations, creating charge density and flux of charge, an electric current that varies with time, or frequency.

Most systems cannot be described by a single harmonic oscillator. Combinations of harmonic oscillators have more complex properties. First consider a parallel combination of oscillators, in which each oscillator is independent of the others and depends on fields (and everything else) exactly as a single oscillator does. Combinations of independent harmonic elements will have a distribution of time dependent properties that is more or less the sum of each element if the forces on one element are independent of the location and parameters of the other elements. For example, if one measures the total current of elements in parallel, the current will be the sum of the distribution of currents of each element. The properties of each element and of the distribution of elements will however produce complex time dependent currents not describable by the properties of a single harmonic oscillator, or (in the frequency domain) by a single dielectric constant. Indeed, if these harmonic oscillators are coupled, as they usually are, they can produce one of the most chaotic systems known to mathematics or science.

Most systems contain elements that are not independent. Each element (of a mass with charge on a spring) will exert force on its neighbors and the properties of the whole system will not be the sum of the individual (isolated) elements. These interactions cannot be described by a single potential field that is the same for all the independent oscillators. A potential field acting on one element will depend on the properties of the other elements and so the function describing the potential field will be different for each element. Even the potential field produced by a perturbation (say a perturbation applied by electrodes at the boundaries as experiments are usually done) will depend on the properties of other elements. The perturbing potential will create an applied field that will move each element and that change in location will change the force on every other element. The applied field acting on one element will not be a function of just the perturbation potential. Combinations of interacting masses (with charge) are likely to have properties that differ qualitatively from the properties of individual (isolated) elements or a distribution of isolated elements. Hence, this system is now an interacting many-body system, and becomes one of the most difficult problems to solve in either classical or quantum physics or chemistry.

The harmonic oscillator discussed by Hall and Heck is not an artificial example. The classical harmonic oscillator is used throughout theoretical physics from Planck's treatment of quantized light, arising from an ensemble of such oscillators, even in studies of the quantum vacuum (Milonni 2013) through quantum mechanics (e.g., (McIntyre, Manogue et al. 2013). It is not an exaggeration to say that study of the harmonic oscillator is the starting point of most of many body physics (Ch.1 of (Mahan 1993)).

Chemical compounds are a hierarchy of partially coupled charged oscillators. Each bond oscillates as electric fields change. And bonds are electrical objects (distributions of electrons) linking atoms that usually have significant charge. Groups of atoms together form units ('moieties' is a name commonly used) that move together, more or less—more rather than less in many important cases. These compounds form a hierarchy of nested oscillators, one building on another, that make a compound pendulum look simple. Compound pendulums have remarkably complex motions. Chemical compounds consisting of a hierarchy of nested charged oscillators will clearly not be describable by a single harmonic oscillator, let alone a single dielectric coefficient, even if they are in solids, or in an ideal gas.

In liquids, polarization is more complex and hard to describe in a general way because liquids are far more deformable than solids. In liquids, matter and charge move in ways rarely found in solids. Long distance flows of mass and charge driven by non-uniform boundary conditions are characteristic of liquids and not of solids, although of course fields of quasi-particles in solids (like holes and electrons of semiconductors) flow much like ionic liquids. Movements of charge are often driven by nonelectric forces like diffusion or convection. Description of polarization in such systems must include the field equations of diffusion or convection and their coupling to the field equations of electricity, along with the boundary conditions that are an integral unavoidable part of the definition of such fields that can have important practical consequences (Mertens and Weeks 2016).

Many experiments have shown the complexity of polarization in liquids. Polarization has been studied extensively in the ionic solutions derived from sea water in which life occurs and in which much of chemical experimentation is performed. Experiments show that polarization currents cannot be approximated by a dielectric coefficient that is a real positive constant over any reasonable range of conditions or scales (Oncley 1942, Nee and Zwanzig 1970, Macdonald 1992, Barthel, Buchner et al. 1995, Barthel, Krienke et al. 1998a, Barthel, Krienke et al. 1998b, Buchner and Barthel 2001, Kremer and Schönhals 2003, Oncley 2003, Barsoukov and Macdonald 2005). The magnitude of the effective dielectric coefficient (as usually defined in experiments in the frequency domain) varies by a factor of 40× and the current and voltage are not even approximately in phase: delays abound and the delays depend dramatically on frequency, concentration of ion, and types of ions present. (A glance of the extensive data in Barthel (Barthel, Buchner et al. 1995) is instructive.) Worse, under such circumstances, polarization current must be described by convolution-type integrals[4] that do not easily fit into the formalism of Maxwell's constitutive **D** field (Abraham and Becker 1932, Purcell and Morin 2013) that depends on a constant dielectric coefficient, a single real number.

**1.4 Historical Note.** Readers may jump over this note without losing the general trend of the paper, if they wish.

Despite the overwhelming experimental evidence, and theoretical understanding of complex polarization, the complexity is not recognized in many areas of science and most treatments of electrodynamics and Maxwell's equations. The implications of complex dielectric behavior for transient properties is not apparent in the classical approach focused on sinusoids ***at one frequency.*** In the present world, we are interested in atomic motions which are nearly white noise, more or less the sum of sinusoids of all frequencies, with an extraordinarily large numbers of reversal of directions in even $10^{-15}$ sec and so the simplifications of sinusoidal analysis at one frequency are not of much help. We hope the following discussion makes clear how confusion arose and so makes it easier to move towards reality and whatever clarity it permits.

Textbooks have used a single time independent dielectric coefficient (a real positive number) since at least 1893, as described in histories (Holton 1967, Mehra 2001, Arthur 2013) and by physicist and textbook authors Max Abraham and Richard Becker whose early texts (Abraham and Becker 1932, Becker and Sauter 1964; with editions going back to Abraham-Föppl, 1905) were the foundation for so many others. Textbook treatments of dielectrics tend to be built on each other, rather than on the actually observed properties of real materials.

The appropriate mathematical generalization for variable dielectric coefficients is not found in the references cited. They almost all use a frequency dependent (i.e., variable) dielectric coefficient (that is a complex number with real and imaginary parts, not a real number or real constant, but rather a complex variable) and concentrate on the frequency domain case. Analysis begins with constant dielectric coefficients in the differential equations and then turns that constant into a variable in the use of the solution of those equations. Whatever help this may be in dealing with sinusoids of one frequency disappears when dealing with transient responses even to step functions, let alone to (nearly) white noise of atomic motion. At best one must perform inverse Laplace transforms of considerable difficulty to extend to the time domain. These nearly always lead to complex convolutions in expressions that do not fit comfortably into the usual **D** field formulation of Maxwell's equations. Often the inverse Laplace transforms cannot be performed because the system is nonlinear or the mathematics is too difficult. In biological systems and condensed phases, the system is nearly always driven by forces not included in Maxwell's equations, so a much more general treatment is needed, that benefits from variational methods designed to combine different forces consistently.

The mathematically obvious needs to be restated because all scientists are human. It is only human to try to extend ideas, to see how far we can go, to see what happens if we stretch a constant into a variable. In fact, one of the standard methods of solving differential equations presumably arose from an attempt to stretch constants into variables. It is called 'variation of constants' or 'variation of parameters' for that reason (Tenenbaum and Pollard 1963, Arnol′d 2012). This method produces terms, however, that are ***not*** present in the solution of equations with constant inhomogeneous terms. The variation of parameters produces a different form of the solution of the differential equations. If the constants in the solution were turned into variables, these additional terms would not be present and so the 'solution' involving only the terms of the original differential equation would no longer satisfy the differential equation (with variable coefficients).

[4] Such convolutions occur throughout physics. They commonly arise in systems that are far from equilibrium, possess several different "time constants" and so cannot easily be written as a scalar Markov process. (Karlin and Taylor 1975,Schuss, 1980, Schuss 2009).

The full treatment 'variation of constants' is needed to solve differential equations because mathematics does not allow self-contradiction. A constant in part of a derivation must remain a constant in the whole derivation, including the result of the derivation. A constant cannot become a variable. This statement is obvious, but dielectric constants (real positive numbers) have been turned into variables (complex frequency dependent variables) as common practice, throughout the literature of dielectric coefficients for more than a century. And so this and the surrounding paragraphs are needed, we fear, if we are to be absolutely explicit and convincing, so we can change a common practice so deeply embedded in our history.

If one assumes a constant dielectric coefficient in a differential equation, and solves the equation with that assumption, it is incorrect mathematics to extend the solution into a new formula by allowing a parameter to become a variable. Imagine that the variable dielectric coefficient were included in a second generalized differential equation. That revised equation would have a different solution from the extended formula. A formula that is an extension of the solution (using a variable dielectric coefficient) will not satisfy the generalized differential equation that includes a variable dielectric coefficient. The solution to the differential equations are different formulae.

### 1.4 Structure of the paper

In section 2, we provide an atomic scale discussion (at a fundamental level) about the intrinsic origins of the particle and displacement currents. We deduce such currents from the trajectories of particles. We also show in this section that all developments in terms of trajectories are fully compatible with quantum phenomena. In section 3, we abandon the atomic level of description and develop macroscopic Maxwell equations when a spatial average of the atomic magnitudes is warranted. There, we present the macroscopic particle and displacement currents in idealized systems. Section 4 shows that a quite different approach is needed to deal with realistic systems, but that approach can provide crucial results. Conservation of current is a universal law that can be derived independent of the polarization properties of matter, for example. Finally, we provide some concluding remarks in section 5.

## 2. ATOMISTIC PARTICLES AND DISPLACEMENT CURRENTS

Ignoring the structure of the nucleus of atoms (which is far from the scope of the present work), we can consider that electrons, atoms (or ions or molecules) are the fundamental particles of our system. We will discuss the particle current and the displacement current assigning a trajectory to each of these particles. We will also show that such trajectory-based understanding of the currents is also perfectly compatible for all (non-relativistic) quantum phenomena. Hence, no real change in the understanding of the role of the electric fields occurs as we move from classical to quantum treatments.

## 2.1 The Particle Current

We consider a general system of $N$ particles. Each particle has a mass $m_i$ and a charge $q_i$ (the charge $q_i$ can be a positive or negative number, or even zero for neutral particles, but the mass $m_i$ is always a positive number). Each particle is defined by a trajectory $\mathbf{x}_i(t)$ in three dimensional space. We will use normal symbols to define scalar values and bold symbols for vectors in this section. A set of $N$ trajectories $\{\boldsymbol{x}_i[t]\}$ with $i = 1, \dots, N$ provides a description of our system. The charge density of such system can be defined as:

$$\rho_Q \equiv \rho_Q(\boldsymbol{x}, t) = \sum_{i=1}^{N} q_i \delta(\boldsymbol{x} - \boldsymbol{x}_i[t]) \quad (2)$$

where $\delta(\boldsymbol{x})$ is the Dirac delta function that specifies the position at which the particle is located. In order to simplify the notation, the dependence on $\boldsymbol{x}$ and $t$ will not be explicitly indicated, unless necessary. Similarly, we will use $\boldsymbol{x}_i \equiv \boldsymbol{x}_i[t]$ without writing the explicit time dependence. The time dependence of such charge density, because of the movements of the particles, can be evaluated as:

$$\frac{\partial \rho_Q}{\partial t} = \frac{\partial}{\partial t} \sum_{i=1}^{N} q_i \delta(\boldsymbol{x} - \boldsymbol{x}_i) = \sum_{i=1}^{N} q_i \boldsymbol{\nabla} \delta(\boldsymbol{x} - \boldsymbol{x}_i) \cdot \left(-\frac{d\boldsymbol{x}_i}{dt}\right)$$
$$= -\sum_{i=1}^{N} q_i \boldsymbol{v}_i \cdot \boldsymbol{\nabla} \delta(\boldsymbol{x} - \boldsymbol{x}_i) = -\nabla \cdot \left(\sum_{i=1}^{N} q_i \boldsymbol{v}_i \delta(\boldsymbol{x} - \boldsymbol{x}_i)\right) \quad (3)$$

where $\nabla$ is the divergence operator acting on $\boldsymbol{x}$ and $\boldsymbol{v}_i \equiv \boldsymbol{v}_i[t] = d\boldsymbol{x}_i[t]/dt$ is the velocity of particle $i$ in the three dimensional space. We define now the particle current density of the $N$ particles as,

$$\boldsymbol{j}_Q = \sum_{i=1}^{N} q_i \boldsymbol{v}_i \delta(\boldsymbol{x} - \boldsymbol{x}_i) \quad . \quad (4)$$

The subindex $Q$ just indicates that we are dealing with a flux of particles at position $\boldsymbol{x}$ and time $t$ as indicated in table 1. Then, Eq. (3) can be rewritten in the form of the well-known local conservation law:

$$\frac{\partial}{\partial t} \rho_Q + \nabla \cdot \boldsymbol{j}_Q = 0 \quad . \quad (5)$$

This law is satisfied by all systems that are composed of particles with a real mass, whether at a classical or quantum level, and with or without charge. We notice that Eq. (5) forbids, for example, any model where a particle disappears (instantaneously, without delay) from its original position and reappears (immediately, without delay, at the same time it disappeared) at another point far away from its original location. From the definition of the particle current density in Eq. (4), we see that a large particle current can imply either many particles with small velocity or few particles with large velocity. This variety of dynamics is captured in most hydrodynamic models of transport in chemistry and biology.

## 2.2 Displacement Current

When we are considering a system with charged particles, these particles must satisfy the requirements imposed by the interactions due to the charge. The charge and the particle current densities due to the motion of that charge have to satisfy Maxwell's laws. The first of these we call Gauss's law:

$$\varepsilon_0 \nabla \cdot \boldsymbol{e} = \rho_Q \tag{6.1}$$

where $\boldsymbol{e} \equiv \boldsymbol{e}(\boldsymbol{x}, t)$ is the *atomic scale* electric field generated at the position $\boldsymbol{x}$ and time $t$ by the set of particles whose positions are $\{\boldsymbol{x}_i[t]\}$. We will use capital letters later for the *macroscopic* fields. The term $\varepsilon_0$ is the permittivity of free space (also defined as the vacuum permittivity, and introduced in the previous section). In addition, the following equations also have to be satisfied by our system of charged particles:

$$\nabla \cdot \mathbf{b} = 0 \tag{6.2}$$

$$\boldsymbol{\nabla} \times \mathbf{e} + \frac{\partial \mathbf{b}}{\partial t} = 0 \tag{6.3}$$

where $\boldsymbol{b}(\boldsymbol{x}, t)$ is the atomic scale magnetic field. Finally, the fourth Maxwell equation is Ampere's law with Maxwell's modification:

$$\frac{\boldsymbol{\nabla} \times \mathbf{b}}{\mu_0} = \boldsymbol{j}_Q + \varepsilon_0 \frac{\partial \mathbf{e}}{\partial t} \tag{6.4}$$

where $\mu_0$ is commonly called the vacuum permeability, permeability of free space or magnetic constant. The speed of light in free space $c_0$ can be defined as $c_0 = 1/\sqrt{\mu_0 \varepsilon_0}$ and is remarkably determined by electrical and magnetic properties that can be measured entirely independent of light.

By introducing Eq. (6.1) into Eq. (5) we get the following result:

$$\frac{\partial}{\partial t}(\varepsilon_0 \nabla \cdot \boldsymbol{e}) + \nabla \cdot \boldsymbol{j}_Q = \nabla \cdot \left(\varepsilon_0 \frac{\partial \boldsymbol{e}}{\partial t} + \boldsymbol{j}_Q\right) = 0 \tag{7}$$

Identical results can be obtained from the divergence of (6.4). The first term on the right hand side of Eq. (7) is a new type of current density related to the time-dependence of the electric field, and which we have introduced already in eqn. (1). This term is non-zero at the position $\boldsymbol{x}$ and time $t$ even when there is no particle there. The new current term arises either from conservation law (5) and the electrostatic equation (6.1) or from the magnetic field equation (6.4). Both derivations give the same result. Eq. (5) establishes a local conservation of particles, while Eq. (7) establishes a local conservation of the total current.

In order to understand the implications of Eq. (7) in the description of the dynamics of a system of charged particles, let us consider a volume $\Omega$ limited by a closed surface $S$. The volume is totally arbitrary and can include all the particles, some of them, or none at all, just by defining the volume itself. Then, by applying the divergence theorem (or Gauss's theorem),(Schey and Schey 2005) we get the result:

$$\int_{\Omega} \nabla \cdot \left(\varepsilon_0 \frac{\partial \boldsymbol{e}}{\partial t} + \boldsymbol{j}_Q\right) d^3x = \int_{S} \left(\varepsilon_0 \frac{\partial \boldsymbol{e}}{\partial t} + \boldsymbol{j}_Q\right) \cdot d\boldsymbol{s} = 0 \tag{8}$$

with $d^3x$ a volume differential and $d\boldsymbol{s}$ the differential surface which is a vector locally perpendicular (pointing outwards) to the $S$ surface. From now on, we distinguish between current density and current itself, contrary to the simplification in Section 1.2. If we assume, for example, that the volume $\Omega$ is a parallelepiped with a closed surface $S = \{S_1, S_2, \ldots, S_6\}$, then, we get:

$$\sum_{i=1}^{6} \int_{S_i} \left(\varepsilon_0 \frac{\partial \boldsymbol{e}}{\partial t} + \boldsymbol{j}_Q\right) \cdot d\boldsymbol{s_i} = \sum_{i=1}^{6} I_i(t) = 0 \tag{9}$$

where we use the definition of total current following expressions (1b) in subsection 1.2 as:

$$I_i(t) = I_{i,Q}(t) + I_{i,d}(t); \tag{9.1}$$

$$I_{i,D}(t) = \int_{S_i} \varepsilon_0 \frac{\partial \boldsymbol{e}(\boldsymbol{x},t)}{\partial t} \cdot d\boldsymbol{s_i} \tag{11}$$

$$I_{i,Q}(t) = \int_{S_i} \boldsymbol{j}_Q(\boldsymbol{x},t) \cdot d\boldsymbol{s_i} \tag{10}$$

where we have defined the displacement and particle current in general, and rewritten eq. (1) which was written for a constant lateral area.

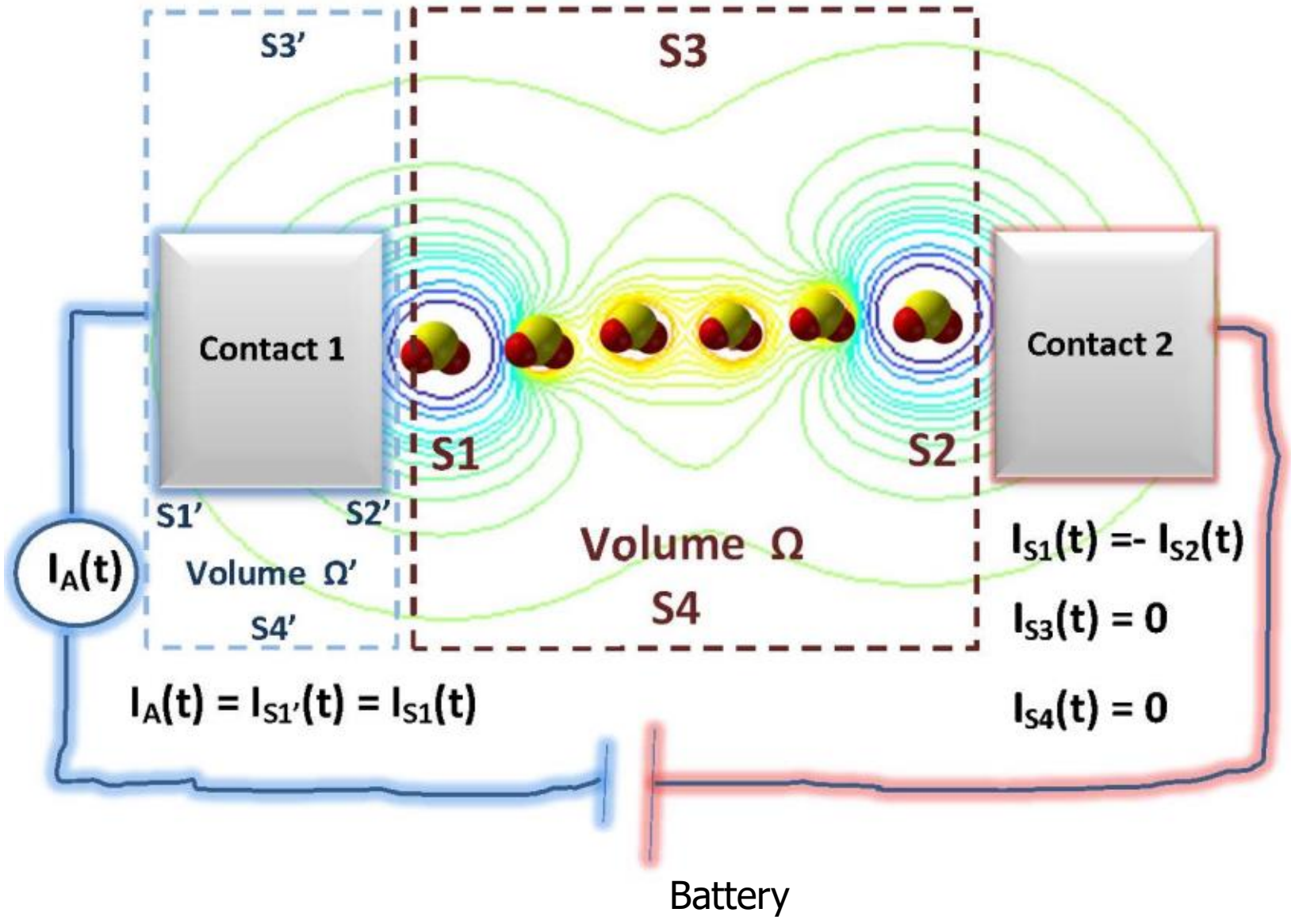

*Fig. 1: A two terminal device with a correct selection of the simulation box $\Omega$ that allows a correct computation of the flux of particles and the electric flux on S1, $I_{S1}(t)$, so that it coincides with the measured current in the ammeter, i.e. $I_{S1}(t) = I_A(t)$.*

The conservation of the total current in Eq. (7) can be illustrated with the 2D example in Fig. 1. Particles move through each of the surfaces $S_1$ and $S_2$. Such a transport of particles generates an electric field everywhere. The intensity of the electric field is larger close to the particles and tends to become negligible at locations far from where the particles are located. Therefore, we can assume that in the side surfaces ($S_3$ and $S_4$ in Fig. 1), there is no particle or displacement current. Then, the volume Ω behaves as a two terminal device (Tuttle 1958, Weinberg 1975). Note the two terminal device can be a transmission line (Ghausi and Kelly 1968) described by partial differential equations of the telegrapher type. These can be exactly described by two port theory of electrical networks and simple analytical expressions involving the usual hyperbolic trigonometric functions. The two port theory of transmission lines provides an interesting link between the engineering world of electrical networks and the mathematical world of field equations which deserves more investigation.[5]

Note that the condition in Eq. (5) can be rewritten here as $I_1(t) = -I_2(t)$. The total current entering into Ω through $S_1$ is equal at every instant of time to the current leaving it through $S_2$. This is true even at the particular moment when a particle leaves through $S_1$ , but no other particle enters through $S_2$. In that moment, continuity of current requires a change in the physical nature of current. The miracle of Maxwell's equations is that they apply no matter what the physical nature of current, or to say the same thing a different way, they produce the exact displacement current needed to guarantee continuity of current at every time no matter what physics governs the flux of charges.

**Electricity is different** from other forces in this respect. Other forces do not have an equivalent of vacuum displacement current $\varepsilon_0(\partial e(\boldsymbol{x},t)/\partial t)$ to enforce exact continuity of (their equivalent of) total current under all conditions, at all times, and in all locations of a series circuit.

The difference between particle current and total current is the displacement current. The equivalence between the two currents moving through the surfaces holds for the total current. The particle currents are not equal, nor are the displacement currents, only the total currents.

If we add another volume *Ω'* at the left side of the original one (see fig. 1), we may then conclude that $I_{S1}(t) = I_{S1'}(t)$. In particular, the total current measured in the ammeter of fig. 1 is equal to the total current computed on the surface of the original volume *Ω* , that is $I_{S1}(t) = I_A(t)$. Again, this argument holds only for the total current and not for only the particle current by itself, nor for the displacement current by itself.

An even more surprising example of the relevance of Eq. (7) appears in a two-terminal capacitor. In the capacitor, there is transport of total current along all points of the capacitor without any passage of particles through the volume of the capacitor. There, the external particle current is matched by the internal displacement current. If we consider another example where three surfaces have non-negligible total current, as in a transistor, then we get a three terminal device with a conservation law for the total current written as $I_1(t) + I_2(t) + I_3(t) = 0$.

[5] Inverse problems of network synthesis have been analyzed with great success, exploiting the theory of complex variables. In particular, ill-posedness produced by structural redundancy—parallel resistors—has been separated from other parasitic sensitivity (not enough data). It would be interesting to use the two port theory of transmission lines to try to extend this separation of types of ill-posedness to the inverse theory of partial differential equations in general.

In fact, if one considers a series arrangement of typical laboratory devices connected by wires, devices like resistors, capacitors, batteries and diodes (Fig. 2 of (Eisenberg 2016c) it is clear that currents in each device arise in very different ways, that vary a great deal with time, yet the current in each device is exactly equal at all times, no matter what the physical origin of the current. The displacement current arranges itself to satisfy Maxwell's equations and make this happen—eq. 4 of (Eisenberg 2016b) shows one way this can happen—in all devices at all times, no matter how the currents arise from the motion of charged particles.

## 2.3 Particle and Displacement Currents in Quantum Systems

In the previous sections, our discussions about particle and displacement currents may have been viewed as applicable only to classical systems (Zapolsky 1987, Arthur 2008, Selvan 2009, Arthur 2013). This is not true. All of our discussion about the two components of the total currents can be directly applied to (non-relativistic) quantum systems. In classical systems, the particle motion arises from the Hamiltonian, or total energy. This is still true in quantum systems although new quantum potentials/forces supplement the classical Hamiltonian (Kennard 1928, Bohm 1951). The trajectory of each quantum particle is associated with a quantum (Bohm) trajectory, $\boldsymbol{x_i}[t]$. Certainly, we could try a more orthodox description of particle and displacement currents in quantum systems without trajectories, but treatment of current and displacement current becomes more difficult, in our view.

We believe that the trajectory-based description of quantum mechanics (which we will explain here) provides a much simpler treatment of particle and displacement currents, even almost trivially so, than the orthodox one. After all the orthodox approach must consider the 'measurement problem' of orthodox quantum mechanics for both particle current and displacement current. And however one thinks of measurement in orthodox quantum mechanics, one must admit that it is not simple. The Bohm treatment is simpler because the measurement problem does not require explicit discussion beyond the definition of the treatment itself (Oriols and Mompart 2012, Dürr, Goldstein et al. 2013, Benseny, Albareda et al. 2014)

Yet we admit that explanations of quantum phenomena in terms of quantum trajectories and waves are not as popular as explanations with waves alone. Hence, we first give a brief discussion of the empirical equivalence between different quantum theories as they are pertinent here. (Readers may jump over this history, to eq. (12), without losing the general trend of the paper, if they wish).

The Copenhagen interpretation (Born, Heisenberg et al. 1925, Born and Jordan 1925, Born 1926), Bohm mechanics (de Broglie 1925, Bohm 1951), consistent histories (Griffiths 1984, Omnes 1988, Gell-Mann and Hartle 1990) , and instantaneous collapse theories (Ghirardi, Rimini et al. 1986) are just a few of the various interpretations of quantum phenomena that give identical empirical results for all experiments, while being different ontological theories. To better understand the differences between empirical and ontological planes of a theory, we briefly enter into the discussion of what is a physical theory. Kant was the first to divide scientific knowledge into three parts: appearance, reality and theory (Herbert 1987). Appearance is the content of our sensory experience of natural phenomena, i.e. the empirical outcome of an experiment. It might be called the estimator of reality if we used the language of statistical inference and estimation theory (Sorenson 1980, Efron 1982, Stengel 1994, Tarantola 2005), where the difference between estimators and reality is a central subject, of great practical importance. Reality is what lies behind all natural phenomena. A theory is a human model that tries to mirror both appearance and reality. The particular reality invoked (e.g., predicted) by a theory is referred as the ontology of the theory. Empiricists believe only in experimental outcomes (what Kant called appearance) and refuse to speculate about what deeper reality the theory implies. On the other hand, realists believe that good physical theories explain, or at least provide clues about, the reality of our comprehensible world.

The Copenhagen interpretation, for example, assumes that the reality of a quantum system is somehow undefined until a measurement on the system is done (Heisenberg 1925). The wave function solution of the Schrödinger equation is not viewed as providing a description of the reality of an individual experiment, but only provides a compact description of the probabilities associated to all possible experiments/realities (Heisenberg 1927). According to the Copenhagen interpretation, one particle, for example an electron, is sometimes a wave and other times (when a position measurement is done) is a particle. The difficulties in properly understanding how a unique quantum entity can be a wave or a particle reality, and change between the two when a collapse occurs, just shows the difficulty in accepting the (somehow schizophrenic) ontology of the Copenhagen interpretation.

As we have said, there are other quantum interpretations, which also have total agreement with experimental results, while invoking a different understanding (ontology) of reality. In particular Bohm mechanics explains, in a trivial way, the dual role of an electron as *both* a wave and as a point-particle (the fact that a light photon was required to have this duality was known as early as 1909 (Taylor 1909). The theory uses two objects, one wave and one point-particle, to describe just one electron. Then, the wave-particle duality is understood with Bohm mechanics as easily as we understand a classical (point-particle) electron which is being guided by an (wave) electric field. Moreover, this interpretation allows us to clearly identify trajectories which are quite similar to those in classical physics. As mentioned above, the particles in these trajectories obey Hamiltonian mechanics, just as classical particles do, but in addition respond to additional quantum potentials (Kennard 1928).

The first element in the Bohm theory for a describing the system of $N$ particles mentioned previously is the wave function $\Psi \equiv \Psi(\boldsymbol{x}_1, \ldots, \boldsymbol{x}_N, t)$ in the multi-dimensional *configuration* space, and which is a solution of the many-particle Schrödinger equation:

$$i\hbar \frac{\partial \Psi}{\partial t} = \left\{ -\sum_{i=1}^{N} \frac{\hbar^2 \nabla_i^2}{2m_i} + u \right\} \Psi \tag{12}$$

where $\nabla_i^2$ is the Laplacian operator acting on $\boldsymbol{x}_i$. The potential energy $u \equiv \mathrm{u}(\boldsymbol{x}_1, \ldots, \boldsymbol{x}_N, t)$ reflects the interaction between the $N$ particles among themselves as well as any external potentials. For example, it can include the Coulomb interaction among particles. We emphasize that the wave function is defined in the configuration space, not in the ordinary three dimensional real space—the configuration space has three dimensions for each particle so that the total dimension is $3N$. However, our intuition is developed for the three dimensional physical space and this explains why some quantum phenomena like non-local correlation between distant particles (what Albert Einstein defines as "spooky action at a distance") becomes counter-intuitive (and, in fact, unnecessary in a realist viewpoint (Ferry, 2018)). Our concept of distance between two objects is valid for a three dimensional physical space, but it loses its meaning in the $3N$ dimensional configuration space. We notice that scalar potential energy $u(\boldsymbol{x}_1, \dots, \boldsymbol{x}_N, t)$ in (12) is also a non-local potential and is also defined in this huge $3N$ dimensional configuration space. Neglecting relativistic effects, one reasonable solution for the potential is:

$$\mathrm{u}(\boldsymbol{x}_1, .., \boldsymbol{x}_N, t) = \sum_{i=1}^{N} \sum_{j>i}^{N} \frac{1}{4\pi\epsilon_0} \frac{q_i q_j}{|x_i - x_j|} \tag{13}$$

In principle, one can also include the magnetic interaction among charged particles in Eq. (12) by adding the vector potential in the definition of the momentum operator.

We have assumed a closed quantum system in the sense that the set of $N$ particles are properly described by a pure state, not by a reduced density matrix. Open systems can be modelled by a closed one by adding all the rest of the particles of the environment or by connecting with appropriate boundary conditions, and other field equations, as appropriate. Indeed, much of condensed matter physics, engineering, and biology is devoted to open systems and we spend much time on open systems later in this paper.

At this point, we notice that Eq. (12) contains a local conservation law for the quantum probability density $\rho_q = |\Psi|^2$:

$$\frac{\mathrm{d}\rho_q}{dt} + \sum_{i=1}^{N} \nabla_i \cdot j_i = 0 \tag{14}$$

where $j_i \equiv j_i(\boldsymbol{x}_1, \dots, \boldsymbol{x}_N, t)$ is the (ensemble value of the) quantum current density and $\nabla_i$ the divergence vector on the $\boldsymbol{x}_i$ position (Landau and Lifshitz 1958). We have used Eq. (3) and (4), written with trajectories to deduce a conservation law in (14). The inverse reasoning has been used by many scientists to suggest that quantum trajectories are, in fact, hidden in Eq. (14) or that a trajectory-based interpretation of quantum phenomena is possible within Eq. (12). Many scientists have noticed the analogy with Langevin trajectories and Fokker Planck equations describing the density of those trajectories.(Karlin and Taylor 1975, Karlin and Taylor 1981, Schuss 2009)

The second element of the Bohm theory when describing a $j$-experiment is a set of well-defined trajectories in the normal three dimensional physical space $\{\boldsymbol{x}_1^j[t], \dots, \boldsymbol{x}_N^j[t]\}$. The superindex $j$ specifies that the Bohm definition of the quantum state refers only to one $j$-experiment. The velocity of each particle for $k = 1, .., N$ is defined from the wave function as:

$$\mathbf{v}_{\boldsymbol{k}}^{\boldsymbol{j}}[t] = \frac{d\mathbf{x}_{\boldsymbol{k}}^{\boldsymbol{j}}[t]}{dt} = \frac{J_k\left(x_1^j[t], .., x_N^j[t], t\right)}{\left|\Psi\left(x_1^j[t], .., x_N^j[t], t\right)\right|^2} \tag{15}$$

By time-integrating Eq. (12), the trajectory of each particle can be computed trivially as:

$$\mathbf{x}_k^j[t] = \mathbf{x}_k^j[0] + \int_0^t \mathbf{v}_k^j[t']dt' \tag{16}$$

To get the exact trajectory, we have to specify the initial position of each particle in the experiment. Contrary to classical mechanics (where the measurement of the initial positions of a system is considered unproblematic), the initial position of the Bohm particles cannot be measured (unless the many particle initial wave function is close to a delta function for each position). In general, in quantum mechanics, only probabilities of the different outputs of experiments can be predicted. There is an unavoidable uncertainty in quantum phenomena. ***In the Bohm theory, the quantum uncertainty is implicit in the uncertainty of the initial positions.*** Experiments are modelled many times, j= $1, \ldots., M \to \infty$ , with the same wave function $\Psi(\boldsymbol{x_1}, \ldots, \boldsymbol{x_N}, \boldsymbol{y}, t)$, but with different initial positions for each set o $N$ trajectories. The probability distribution of the set of trajectories in different experiments is given by

$$|\Psi(\boldsymbol{x_1}, \ldots, \boldsymbol{x_N}, t)|^2 = \frac{1}{M}\sum_{j=1}^{M} \delta\left(\boldsymbol{x_1} - \boldsymbol{x}_1^j[t]\right) \ldots \delta(\boldsymbol{x_N} - \boldsymbol{x}_N^j[t]) \tag{17}$$

The construction of the Bohm trajectories through Eqs. (15)-(16) ensures that if a large ensemble of experiments $j = 1, \ldots., M \to \infty$ with $N$ trajectories $\left\{\boldsymbol{x}_1^j[t], .., \boldsymbol{x}_N^j[t]\right\}$ in each experiment are selected in agreement with (17) at a particular time $t = 0$, then, the distribution $|\Psi(\boldsymbol{x_1}, \ldots, \boldsymbol{x_N}, t)|^2$ will be satisfied by those set of trajectories at any other time. The reason why the Bohm and Copenhagen theories are empirically equivalent is due to this equivariance condition implicit in (17) (Oriols and Mompart 2012, Dürr, Goldstein et al. 2013, Benseny, Albareda et al. 2014).

Contrary to the wave function that 'lives' in the $3N$ dimensional configuration space, the Bohm trajectories $\left\{\boldsymbol{x}_1^j[t], \ldots, \boldsymbol{x}_N^j[t]\right\}$ in a single experiment 'live' without problem in the normal three dimensional physical space. Therefore, in a single experiment, the charge density at the point $\boldsymbol{x}$ in the physical space due to the other particles $\left\{\boldsymbol{x}_1^j[t], \ldots, \boldsymbol{x}_N^j[t]\right\}$ can be trivially defined as:

$$\rho_Q^j(\boldsymbol{x}, t) = \sum_{i=1}^{N} q_i \delta\left(\boldsymbol{x} - \boldsymbol{x}_i^j[t]\right) \tag{18}$$

where the superindex $j$ means that this charge density corresponds only to the $j$-experiment. In another experiment, the charge can be different due to the intrinsic quantum uncertainty in the selection of the initial positions. From $\rho_Q^j(\boldsymbol{x}, t)$ and the Poisson equation, we can define the potential $\mathrm{v}^j(\boldsymbol{x}, t)$ as the potential created at the point $\boldsymbol{x}$ in the physical space due to the presence of charges at the fixed positions $\left\{\boldsymbol{x}_1^j[t], \ldots, \boldsymbol{x}_N^j[t]\right\}$ as:

$$\nabla^2 \mathrm{v}^j(\boldsymbol{x},t) = -\frac{\rho_Q^j(x,t)}{\varepsilon_0} \tag{19}$$

The boundary conditions in our particular system, where the number of particles $N$ include all relevant particles of the closed system, will be $V^j(\boldsymbol{x} \to \pm\infty, t) = 0,$ which are compatible with the typical Coulomb law. In fact, the solution of (19) gives a potential given by

$$\mathrm{v}^j(\boldsymbol{x},t) = \sum_{i=1}^{N} \frac{1}{4\pi\epsilon_0} \frac{q_j}{\left|x - x_i^j[t]\right|} \tag{20}$$

Once we get this potential, we can compute the electrical field from $\boldsymbol{e^j}(\boldsymbol{x},t) = -\boldsymbol{\nabla}\mathrm{v}^j(\boldsymbol{x},t)$ or from Gauss' law as $\varepsilon_0 \nabla \boldsymbol{e^j}(\boldsymbol{x},t) = \rho_Q^j(\boldsymbol{x},t)$ as mentioned in (6.1). Both expressions give the electric field at the position $\boldsymbol{x}$ due to the $N$ particles as

$$\boldsymbol{e^j}(\boldsymbol{x},t) = \frac{1}{4\pi\epsilon_0} \sum_{i=1}^{N} \frac{q_i}{\left|x - x_i^j[t]\right|^3} (\boldsymbol{x} - \boldsymbol{x}_i^j[t]) \tag{21}$$

Once we know the electrical field at any position $\boldsymbol{x}$, we can compute the displacement current on the points $\boldsymbol{x} \in S_i$ as done in eq. (11). On the other hand, the particle current density of electrons described by Bohm trajectories at the position $\boldsymbol{x}$ can be easily formulated from Eq. (10). It can be easily shown that the ensemble values obtained from eq. (18) are exactly identical to the ensemble values obtained from the Copenhagen interpretation (Albareda, Traversa et al. 2012). ***The fundamental advantage of the Bohm theory is that the total current $I^j(t)$ is well-defined, at any time, with or without discussing its measurement.*** In the present context which is focused on the meaning and properties of 'current' this is a significant advantage over versions of quantum mechanics in which current must involve a whole theory of measurement. The reader is probably aware that scientists do not all use the same quantum theory of measurement.

Another point that requires a clarification is just how we can extract the information $I^j(t)$ from such systems. Such information requires a measurement of the system. In the Bohm theory, the measurement requires the introduction of a pointer (for example the arrow of an analog ammeter) whose position $\boldsymbol{y}$ indicates the value of the measurement of the displacement current. Therefore, we have to introduce a new degree of freedom $\boldsymbol{y}$ in Eq. (14) and also consider the interactions between $\boldsymbol{y}$ and the rest of particles in the Hamiltonian of Eq. (14) so that there is a good correlation between $\boldsymbol{y}$ and $I^j(t)$. Since the degree of freedom $\boldsymbol{y}$ is present in the Schrödinger equation (11), we accept that $\boldsymbol{y}$ is affected by $\{\boldsymbol{x}_1, \ldots, \boldsymbol{x}_N\}$, but we also consider that $\{\boldsymbol{x}_1, \ldots, \boldsymbol{x}_N\}$ are affected by $\boldsymbol{y}$**.** In other words, the evolution of $\{\boldsymbol{x}_1, \ldots, \boldsymbol{x}_N\}$ with or without the ammeter will be different because the solution of (11) will be different. Therefore, the wave function of the quantum systems suffers a back-action due to the measurement. Classically, one accepts (at least theoretically) that one can get information of the particle system without distorting the system. One can imagine an amplifier, for example, with an infinite input impedance that draws no current from is surrounds. In a quantum system the measurement-without-distortion is not possible. It has been demonstrated quite recently by one of the authors that measurement of the displacement current in a quantum system can be considered as a type of weak measurement (Marian, Zanghi et al. 2016). This implies that a good measuring apparatus will provide a value $\boldsymbol{y}^j[t] \approx I^j(t) + \eta(t)$ where $\eta(t)$ is a (very) high frequency noise with ensemble value equal to zero (when integrated over different experiments) and that decays rapidly to zero when time-integrated. In a classical-like language, the physical origin of this extra noise due to the measurement can be attributed to plasmons in the contacts, associated with the displacement current of the weak measurement.

Finally, we emphasize that the quantum reality suggested by each quantum interpretation (ontology) is mainly a relevant topic for those devoted to a realistic understanding of our comprehensible world. Empiricists bother less with the suggested reality as long as the interpretation is empirically correct.

In fact, most scientists are neither realists, nor empiricists; but a mix of both. Many people accept the Copenhagen ontology because it provides a useful method to get practical predictions. The technical advantages in the computation of empirical outcomes is said to compensate somehow the digestive problems implicit in that Copenhagen interpretation of the reality.

For the discussion of the displacement current in this paper, we argue that the Copenhagen interpretation has no technical advantage over the Bohm one, but just the opposite. Thus, for those who like the reality suggested by the Bohm theory, the present description of the particle and displacement current in quantum systems has been found quite simple and intuitive. Those who dislike this Bohm picture of explaining displacement and particle currents in terms of well-defined quantum trajectories can just ignore such reality and use Bohm mechanics as a useful computational tool that helps evaluate and discuss the particle and displacement currents in quantum systems.

If we pursue this subject in more detail, we recognize that the full quantum state (including the active region, the contacts, the batteries, etc.) is computationally inaccessible. A computationally accessible solution deals only with the degrees of freedom of a smaller subsystem, referred as the open system (our active device region), while the other degrees of freedom (the environment) are not explicitly simulated (Breuer and Petruccione 2002). The well-known Lindblad master equation (Lindblad 1976) describes the evolution of the reduced density matrix for Markovian systems (when the role of the environment is highly predictable and memoryless). In the description of the dynamics of quantum systems at the very high frequencies that we are interested in, we can hardly say that the system is Markovian. The orthodox extensions of the Lindblad type of solutions based on the reduced density matrix beyond Markovian dynamics are still challenging. The stochastic Schrödinger equation (SSE) is another technique to deal with non-Markovian systems dynamics with states (Diósi, Gisin et al. 1998, Strunz, Diósi et al. 1999). It is based on the continuous measurement theory that allows the definition of a wave function of the open system conditioned on one monitored value associated with the environment. However, it is well-known that the physical interpretation of the monitored value (for example the measured total current in our case) cannot be given to the solutions of the SSE for non-Markovian systems. It was demonstrated by Wiseman and Gambetta that a SSE-type solution of an open system with a physical interpretation of the monitored value as the output of a continuous measurement has to be based on Bohm mechanics (Gambetta and Wiseman 2002, Gambetta and Wiseman 2003). A practical implementation of this type of computational approach showing the technical advantage of the Bohm approach in some cases is explained in a recent work of one of the authors by using a Bohm conditional wave functions (Oriols 2007, Marian, Zanghi et al. 2016, Colomés, Zhan et al. 2017). A general discussion of the approach to open quantum systems can be found in (Barker and Ferry 1980a, Barker and Ferry 1980b). One such open quantum system coupled to a complex environment is the open "quantum dot" in which coupling to the "dot" is by normal transport, and not by tunnelling. This system illustrates the complexities of the system/environment coupling, and has been the subject of several experimental (Bird *et al*., 1997, 2003) and theoretical reviews (Ferry, Burke et al. 2011, Brunner, Ferry et al. 2012, Ferry, Akis et al. 2015) The Coulomb blockade in ionic channels is closely related to this open quantum system.(Grabert and Devoret 1992, Kaufman, McClintock et al. 2015, Feng, Liu et al. 2016)

## 3. IDEALIZED MACROSCOPIC DESCRIPTION OF THE CURRENTS

As we have already commented, any attempt to describe all $\sim 10^{29}$ fundamental charged or uncharged particles with such an atomic scale dynamical description is generally computationally unfeasible. Therefore, most macroscopic descriptions give up any atomic scale spatial resolution of the discrete particles and deal with a supposedly continuous charge and mass density. From a stochastic viewpoint, the continuous functions are measures of the underlying stochastic processes of atomic motion (Karlin and Taylor 1975, Karlin and Taylor 1981, Schuss 2009), for example, a spatial average. From the scientific point of view, the functions are models of some of the properties of the underlying stochastic processes of atomic motion.

### 3.1 Macroscopic Charge Density and Gauss' Law in Isolated Idealized Systems

The following discussion is of idealized isolated systems that permit spatial averaging. More general open systems are discussed later. We present the idealized equilibrium derivation to connect with the widely read textbook literature (Jackson 1999) and to provide enough detail so others may learn to extend the derivation to the non-equilibrium case relevant to devices and other systems with long-range current flow, driven by (for example) spatially inhomogeneous boundary conditions, with (for example) different potentials at different locations on their boundaries. Temporal averaging is another approach, under intensive study by Chun Liu and associates (Ma, Li et al. 2016a, Ma, Li et al. 2016b).

Here, it will be useful to distinguish between some particles that can be grouped together into small stable entities (like molecules) and other particles that move alone. We assume that there are $i = 1, \dots N_e$ particles moving alone (for example electrons) each one located at $\boldsymbol{x}_i[t]$. We also consider that there are $n = 1, \dots, N_{mol}$ stable entities (molecules) and that each molecule has $i_n$=1,…,$M_n$ particles inside. Therefore, the position of the particles that form the molecule can be written as $\boldsymbol{x}_{i_n}[t] = \boldsymbol{\Delta x}_{i_n}[t] + \boldsymbol{x}_n[t]$ with $\boldsymbol{x}_n[t]$ is the position of the center of mass of the molecule. The charge density in (2) can be written as

$$\rho_p(\boldsymbol{x},t) = \sum_{i=1}^{N_e} q_i \delta(\boldsymbol{x} - \boldsymbol{x}_i[t]) + \sum_{n=1}^{N_{mol}} \rho_n(\boldsymbol{x},t) \tag{22}$$

where $\rho_n(\boldsymbol{x},t) = \sum_{i_n}^{N_n} q_{i_n} \delta\left(\boldsymbol{x} - \Delta\boldsymbol{x}_{i_n}[t] - \boldsymbol{x}_n[t]\right)$ is the charge density of the $n$-*th* molecule. For simplicity, hereafter, since it will be evident that we are talking about charge density, the subindex Q will be avoided.

The macroscopic version of the particle and current densities in idealized systems will be obtained by spatial averaging (Russakoff 1970, Jackson 1999). This type of spatial averaging does not allow the extended effects of finite size particles (Eisenberg 2012, Eisenberg 2013a), for example, and worse, it does not allow the infinite range correlations that occur when spatially nonuniform boundary conditions drive flow. Indeed, it is not clear how to include long range electrical currents that flow from one boundary to another in a spatial distribution function (as they do in the devices of our electronic technology).[6]

It is important to note that any equation for this locally averaged $W(\boldsymbol{x})$ will depend on boundary properties, boundary potential, or charge, and may not visibly depend on current flow at all. Surely the spatial distribution function $W(\boldsymbol{x})$ must vary with current flow if such exists. In general, the distribution function and the fields must be analyzed and computed self-consistently with the various flows.

For an isolated idealized macroscopic system, and any atomic scale magnitude $a(\boldsymbol{x},t)$, such as the electric or magnetic fields, or the charge or particle current densities, we can obtain a continuous magnitude $A(\boldsymbol{x},t) = \langle a(\boldsymbol{x},t)\rangle$ by spatial averaging the atomistic magnitude $a(\boldsymbol{x},t)$ over a localized region, following

$$A(\boldsymbol{x},t) \equiv \langle a(\boldsymbol{x},t)\rangle = \int d^3x'\, W(\boldsymbol{x}')a(\boldsymbol{x}-\boldsymbol{x}',t) \tag{23}$$

where

$$W(\boldsymbol{x}) = Ne^{-\frac{r^2}{R^2}} \tag{23.1}$$

with $r^2 = x^2 + y^2 + z^2$ and $R$ specifies the radius of the small spherical volume over which the spatial average takes place. The value $N$ is a normalization constant. If $R$ is larger than the atomic scale separation between particles, the magnitude $\langle a(\boldsymbol{x},t)\rangle$ becomes a continuous function.

Here we use a spatial distribution function $W(\boldsymbol{x})$ that is inspired by equilibrium analysis of simple systems, akin to a perfect or ideal gas (Rowlinson 1963, Berry, Rice et al. 2000). In systems with extended correlations, any Markovian equation for this locally averaged quantity is inadequate (Jacoboni and Lugli 1989, Hess 1991, Ferry 2000, Singer, Schuss et al. 2004, Vasileska, Goodnick et al. 2010). For example, it is clear that the

[6] Electronic devices are defined by their inputs and outputs and their relationship. Inputs and outputs are at different locations on boundaries of the system: boundary conditions are spatially nonuniform. Most devices also require some locations (usually on boundaries) to be maintained at specified potentials by auxiliary devices called power supplies. Spatially nonuniform boundary potentials drive currents throughout the system that change the properties of the system in useful ways. That is why power supplies are used. The currents driven by the spatially nonuniform boundary potentials satisfy conservation laws and so produce correlations reaching to boundaries. Averaging treatments that do not depend on current cannot easily describe devices that have spatially distinct inputs, outputs, and power supplies.

Gaussian cannot exist adjacent to a hard wall boundary which is impenetrable to the particles. Electrical boundary conditions that define the inputs, outputs, and power supplies of devices are unlikely to have Gaussian distributions nearby. The properties of inputs and outputs are the essential features of devices and so this limitation in the use of Gaussians limits applications.

With the Gaussian approximation, charge densities in Eq. (2) can be spatially averaged from Eq. (23) as

$$\rho(\boldsymbol{x},t) \equiv \langle \rho_p(\boldsymbol{x},t)\rangle = \sum_{i=1}^{N_e} q_i W(\boldsymbol{x}-\boldsymbol{x}_i[t]) + \sum_{n=1}^{N_{mol}} \langle \rho_n(\boldsymbol{x},t)\rangle \tag{24}$$

with the charge of each molecule $\rho_n(\boldsymbol{x},t)$ defined just below (22). Now, since a $\boldsymbol{\Delta x}_{i_n}[t]$ is small in comparison to $\boldsymbol{x}_n[t]$, a Taylor expansion of $\langle \rho_n(\boldsymbol{x},t)\rangle$ around the position $\boldsymbol{x}-\boldsymbol{x}_n$ comes from Taylor expansion of $W(\boldsymbol{x}-\Delta\boldsymbol{x}_{i_n}-\boldsymbol{x}_n)$ as

$$W\big(\boldsymbol{x}-\Delta\boldsymbol{x}_{i_n}-\boldsymbol{x}_n\big) = W(\boldsymbol{x}-\boldsymbol{x}_n) - \Delta\boldsymbol{x}_{i_n}\cdot\nabla W(\boldsymbol{x}-\boldsymbol{x}_n) + \cdots \tag{25}$$

where we have neglected the third (unwritten) term of the Taylor expansion (related to the quadrupole moment). [7]

By putting expression (26) into (24), we can wrote $\rho(\boldsymbol{x},t) \equiv \langle \rho_p(\boldsymbol{x},t)\rangle$ as

$$\begin{aligned} \rho(\boldsymbol{x},t) &= \sum_{i=1}^{N_e} q_i W(\boldsymbol{x}-\boldsymbol{x}_i[t]) \\ &\quad + \sum_{n=1}^{N_{mol}} q_n W(\boldsymbol{x}-\boldsymbol{x}_n[t]) - \sum_{n=1}^{N_{mol}} \boldsymbol{P}_n\cdot\nabla W(\boldsymbol{x}-\boldsymbol{x}_n[t]) + \ldots \\ \rho(\boldsymbol{x},t) &= \sum_{i=1}^{N_e} q_i \delta(\boldsymbol{x}-\boldsymbol{x}_i[t]) \\ &\quad + \sum_{n=1}^{N_{mol}} q_n W(\boldsymbol{x}-\boldsymbol{x}_n[t]) - \boldsymbol{\nabla}\sum_{n=1}^{N_{mol}} \boldsymbol{p}_n W(\boldsymbol{x}-\boldsymbol{x}_n[t]) + \ldots \end{aligned} \tag{26}$$

We have defined the polarization vector of the $n^{th}$ molecule as $\boldsymbol{p}_n \equiv \sum_{i_n=1}^{N_n} q_{i_n}\,\Delta\boldsymbol{x}_{i_n}$ and charge of each molecule as $q_n \equiv \sum_{i_n=1}^{N_n} q_{i_n}$. The macroscopic polarization $\mathbf{P}(\mathbf{x},\mathrm{t})$ is

[7] While this is undoubtedly a reasonable procedure from the physical point of view, it should clearly be understood that these terms may not be an adequate approximation to the Taylor series. There are many independent variables and parameters involved and uniform convergence has not been examined, nor errors of approximation. Evaluating the accuracy of approximations like this is not a mathematical nicety. It is necessary if the approximations are to be used reliably. One must never forget the hundreds or thousands of terms needed in a classical multipole expansion (of Coulomb's law in radial coordinates, for example) when the observation point is close to the source point as it usually is in computations of chemical bonds and molecular dynamics.

$$\mathbf{P}(\mathbf{x},\mathrm{t}) = \sum_{n=1}^{N_{mol}} \boldsymbol{p}_n W(\boldsymbol{x}-\boldsymbol{x}_n[t]) = \sum_{n=1}^{N_{mol}} \langle \boldsymbol{p}_n \delta(\boldsymbol{x}-\boldsymbol{x}_n[t])\rangle \tag{27}$$

Finally, we can rewrite the total charge as

$$\rho(\boldsymbol{x},t) \equiv \langle \rho(\boldsymbol{x},t)\rangle = \langle \rho_{free}(\boldsymbol{x},t)\rangle - \boldsymbol{\nabla}\cdot\mathbf{P}(\mathbf{x},\mathrm{t}) \tag{28}$$

and the Gauss (or first of Maxwell's) equation(s) (6.1) become

$$\varepsilon_0 \nabla\cdot \boldsymbol{E}(\boldsymbol{x},t) = \langle \rho_{free}(\boldsymbol{x},t)\rangle - \boldsymbol{\nabla}\cdot\mathbf{P}(\mathbf{x},\mathrm{t}) \tag{29}$$

where we have defined $\rho_{free}(\boldsymbol{x},t) = \sum_{i=1}^{N_e} q_i \delta(\boldsymbol{x}-\boldsymbol{x}_i[t]) + \sum_{n=1}^{N_{mol}} q_n \delta(\boldsymbol{x}-\boldsymbol{x}_n[t])$. We have defined $\mathbf{E}(\mathbf{x},\mathrm{t}) \equiv \langle \boldsymbol{e}(\boldsymbol{x},t)\rangle$ with the obvious property that $\langle \nabla\cdot \boldsymbol{e}(\boldsymbol{x},t)\rangle = \nabla\cdot\langle \boldsymbol{e}(\boldsymbol{x},t)\rangle$ Then, by defining the electric displacement field as

$$\boldsymbol{D}(\boldsymbol{x},t) = \varepsilon_0 \boldsymbol{E}(\boldsymbol{x},t) + \mathbf{P}(\boldsymbol{x},t) \tag{30}$$

the macroscopic version of the Gauss's law can be rewritten as

$$\boldsymbol{\nabla}\cdot \boldsymbol{D}(\boldsymbol{x},t) = \langle \rho_{free}(\boldsymbol{x},t)\rangle \tag{31}$$

Note that the classical vector field $\boldsymbol{D}$ depends on a constitutive law that does not describe actual experiments on matter. When the classical vector field $\boldsymbol{D}$ is used, polarization is described by a single real number, the dielectric constant $\varepsilon_r$. As we have documented in some detail previously, the polarization of matter cannot be described that way; indeed, the polarization of simple models of matter (as harmonic oscillators) cannot either.

It may be helpful to define a vacuum displacement field

$$\boldsymbol{D_0}(\boldsymbol{x},t) = \varepsilon_0 \boldsymbol{E}(\boldsymbol{x},t) + \boldsymbol{P_0}(\boldsymbol{x},t) \tag{31.1}$$

along with

$$\boldsymbol{\nabla}\cdot \boldsymbol{D_0}(\boldsymbol{x},t) = \langle \rho_{everything}(\boldsymbol{x},t)\rangle = \boldsymbol{\rho}_Q \tag{31.2}$$

The vacuum displacement vector field $\boldsymbol{D_0}$ and the companion polarization $\boldsymbol{P_0}$ field does not involve the properties of matter. It does not involve a constitutive law. These fields are as fundamental and universal as the Maxwell equations themselves (Mansuripur and Zakharian 2009). We call $\rho_{everything}$ by the name $\boldsymbol{\rho}_Q$ later in this paper.

### 3.2 The Macroscopic Current Density and Ampere's Law.

The particle charge densities in Eq. (4) can be spatially averaged from Eq. (23) as

$$\boldsymbol{J_p}(\boldsymbol{x},t) \equiv \langle \boldsymbol{j_p}(\boldsymbol{x},t)\rangle = \sum_{i=1}^{N_e} q_i \boldsymbol{v}_i[t] W(\boldsymbol{x}-\boldsymbol{x}_i[t]) + \sum_{n=1}^{N_{mol}} \langle \boldsymbol{j_n}(\boldsymbol{x},t)\rangle \tag{32}$$

with $\langle \boldsymbol{j_n}(\boldsymbol{x},t)\rangle = \sum_{i_n=1}^{N_n} q_{i_n} \boldsymbol{v}_{i_n}[t] W\left(\boldsymbol{x}-\Delta\boldsymbol{x}_{i_n}[t]-\boldsymbol{x}_n[t]\right)$ which implies a definition of the current of a molecule as

$$\boldsymbol{j_n} = \sum_{i_n=1}^{N_n} q_{i_n} \boldsymbol{v}_{i_n} \, \delta\left(\boldsymbol{x} - \Delta \boldsymbol{x}_{i_n} - \boldsymbol{x}_n\right) \tag{33}$$

Using the same Taylor expansion of $W\left(\boldsymbol{x} - \Delta\boldsymbol{x}_{i_n} - \boldsymbol{x}_n\right)$ in (25) , we can rewrite the spatial average of (33) as

$$\begin{aligned} \langle j_n \rangle = & \sum_{i_n=1}^{N_n} q_{i_n}\left(\boldsymbol{\Delta v}_{i_n} + \boldsymbol{v}_n\right) W(\boldsymbol{x} - \boldsymbol{x}_n) \\ & - \sum_{i_n=1}^{M_n} q_{i_n}\left(\boldsymbol{\Delta v}_{i_n} + \boldsymbol{v}_n\right) \Delta\boldsymbol{x}_{i_n} \cdot \nabla W(\boldsymbol{x} - \boldsymbol{x}_n) + \end{aligned} \tag{34}$$

We have defined the velocity of the center of mass of the molecule and its relative motion as $\boldsymbol{v}_n[t] = d\boldsymbol{x}_n[t]/dt$ and $\boldsymbol{\Delta v}_{i_n}[t] = d\boldsymbol{\Delta x}_{i_n}[t]/dt$. As in the charge density, keeping only the first two terms in the Taylor expansion, we get

$$\begin{aligned} \langle j_n \rangle = & \sum_{i_n=1}^{N_n} q_{i_n} \boldsymbol{v}_n W(\boldsymbol{x} - \boldsymbol{x}_n) + \sum_{i_n=1}^{N_n} q_{i_n} \boldsymbol{\Delta v}_{i_n} W(\boldsymbol{x} - \boldsymbol{x}_n) - \sum_{i_n=1}^{M_n} q_{i_n} \boldsymbol{v}_n \Delta\boldsymbol{x}_{i_n} \cdot \\ & \nabla W(\boldsymbol{x} - \boldsymbol{x}_n) \ - \textstyle\sum_{\mathrm{i}_n=1}^{M_n} q_{i_n} \boldsymbol{\Delta v}_{i_n} \Delta\boldsymbol{x}_{i_n} \cdot \nabla W(\boldsymbol{x} - \boldsymbol{x}_n) \quad \ldots \end{aligned} \tag{35}$$

The first term $\sum_{i_n=1}^{N_n} q_{i_n} \boldsymbol{v}_n W(\boldsymbol{x} - \boldsymbol{x}_n) = \langle q_n \boldsymbol{v}_n \delta(\boldsymbol{x} - \boldsymbol{x}_n)\rangle$ is just the spatial average current of the molecule as if it were a point charge $q_n \equiv \sum_{i_n=1}^{N_n} q_{i_n}$. We notice that the second term gives $\sum_{i_n=1}^{N_n} q_{i_n} \boldsymbol{\Delta v}_{i_n} W(\boldsymbol{x} - \boldsymbol{x}_n) = \frac{\partial}{\partial t}\langle \boldsymbol{p}_n \delta(\boldsymbol{x} - \boldsymbol{x}_n)\rangle + (\boldsymbol{v}_n \cdot \nabla)\langle \boldsymbol{p}_n \delta(\boldsymbol{x} - \boldsymbol{x}_n)\rangle$. The third term can be easily rewritten as $-\sum_{i_n=1}^{M_n} q_{i_n} \boldsymbol{v}_n \Delta\boldsymbol{x}_{i_n} \cdot \nabla W(\boldsymbol{x} - \boldsymbol{x}_n) = -\boldsymbol{v}_n \nabla \cdot \langle \boldsymbol{p}_n \delta(\boldsymbol{x} - \boldsymbol{x}_n)\rangle$ . Neglecting again the fourth order term, we can write the fourth term as $-\sum_{i_n=1}^{M_n} q_{i_n} \boldsymbol{\Delta v}_{i_n} \Delta\boldsymbol{x}_{i_n} \cdot \nabla W(\boldsymbol{x} - \boldsymbol{x}_n) = \nabla W \times \left(\frac{1}{2}\sum_{i_n=1}^{M_n} q_{i_n} \Delta\boldsymbol{x}_{i_n} \times \boldsymbol{\Delta v}_{i_n}\right)$. We define the magnetic dipole moment of the $n$-molecule as

$$\boldsymbol{m}_n = \frac{1}{2}\sum_{i_n=1}^{M_n} q_{i_n} \Delta\boldsymbol{x}_{i_n} \times \boldsymbol{\Delta v}_{i_n} \tag{36}$$

Rewrite the fourth term as $-\sum_{i_n=1}^{M_n} q_{i_n} \boldsymbol{\Delta v}_{i_n} \Delta\boldsymbol{x}_{i_n} \cdot \nabla W(\boldsymbol{x} - \boldsymbol{x}_n) = \nabla \times \langle \boldsymbol{m}_n \delta(\boldsymbol{x} - \boldsymbol{x}_n)\rangle$.

Finally, putting all the terms together, and noting that part of the second term and the whole third term become negligible, we get

$$\boldsymbol{J_p}(\boldsymbol{x}, t) \equiv \langle \boldsymbol{j_p}(\boldsymbol{x}, t)\rangle = \langle j_{free}(\boldsymbol{x}, t)\rangle + \nabla \times \mathbf{M} + \frac{\boldsymbol{\partial}\mathbf{P}(\mathbf{x},\mathrm{t})}{\boldsymbol{\partial t}} \tag{37}$$

Similarly to the definition of the macroscopic polarization $\mathbf{P}(\mathbf{x}, \mathrm{t})$ in Eq. (27), we have defined the macroscopic magnetic dipole moment as

$$M(\mathbf{x}, \mathrm{t}) = \sum_{n=1}^{N_{mol}} \langle \boldsymbol{m}_n \delta(\boldsymbol{x} - \boldsymbol{x}_n[t])\rangle \tag{38}$$

Now, we rewrite the Ampere law in (6.d) as

$$\frac{\nabla\times\langle\mathbf{b}(\boldsymbol{x},t)\rangle}{\mu_0} = \langle j_{free}(\boldsymbol{x},t)\rangle + \nabla\times\mathbf{M} + \frac{\partial\mathbf{P}(\mathbf{x},\mathrm{t})}{\partial\boldsymbol{t}} + \varepsilon_0\frac{\partial\langle\mathbf{e}(\boldsymbol{x},t)\rangle}{\partial t} \tag{39}$$

Using the previous definition $\boldsymbol{D}(\boldsymbol{x},t) = \varepsilon_0\boldsymbol{E}(\boldsymbol{x},t) + P(\boldsymbol{x},t)$ and a new definition of the magnetic field intensity $\boldsymbol{H}(\boldsymbol{x},t) = \frac{\boldsymbol{B}(\boldsymbol{x},t)}{\mu_0} - \nabla\times\mathbf{M}$**,** we arrive at a macroscopic version of the Ampere law in (6.d) as

$$\nabla\times\mathbf{H}(\boldsymbol{x},t) = \langle j_{free}(\boldsymbol{x},t)\rangle + \frac{\partial\mathbf{D}(\mathbf{x},\mathrm{t})}{\partial\boldsymbol{t}} \tag{40}$$

The integration in Eq. (23) depends on $t$ of $a(\boldsymbol{x},t)$ and on the variable $\boldsymbol{x}$**'** but not on the $\boldsymbol{x}$ so it can be easily demonstrate that $\langle\nabla\times\mathbf{b}(\boldsymbol{x},t)\rangle = \nabla\times\langle\mathbf{b}(\boldsymbol{x},t)\rangle = \nabla\times\mathbf{B}(\boldsymbol{x},t)$**.** By the same reasoning $\langle\frac{\partial\mathbf{e}(\mathbf{x},\mathrm{t})}{\partial\boldsymbol{t}}\rangle = \frac{\partial}{\partial\boldsymbol{t}}\langle\mathbf{E}(\mathbf{x},\mathrm{t})\rangle = \frac{\partial\mathbf{E}(\mathbf{x},\mathrm{t})}{\partial\boldsymbol{t}}$**.**

## 3.3 The Macroscopic Particle Conservation Law and the Total Current Density

In Section 3.1 we divided the charge density in Eq. (28) between what we call free charge that includes the electron and molecules (as a point particle) charge $\langle\rho_{free}(x,t)\rangle$ plus the terms $\langle\rho_{notfree}(\boldsymbol{x},t)\rangle = -\nabla\cdot\mathbf{P}(\mathbf{x},\mathrm{t})$. In Section 3.2, we divided the current density in Eq. (37) into two parts, the free current $\langle j_{free}(\boldsymbol{x},t)\rangle$ and $\langle j_{notfree}(\boldsymbol{x},t)\rangle = \nabla\times\mathbf{M} + \frac{\partial\mathbf{P}(\mathbf{x},\mathrm{t})}{\partial\boldsymbol{t}}$. The distinction between free and bound currents is discussed later in this paper where it is found to be of limited use in the study of liquids.

It is interesting to realize that the $notfree$ terms satisfy their own continuity equation

$$\frac{\partial}{\partial t}\langle\rho_{notfree}(\boldsymbol{x},t)\rangle + \nabla\cdot\langle j_{notfree}(\boldsymbol{x},t)\rangle = -\frac{\partial\nabla\cdot\mathbf{P}(\mathbf{x},\mathrm{t})}{\partial t} + \nabla\cdot\left(\nabla\times\mathbf{M} + \frac{\partial\mathbf{P}(\mathbf{x},\mathrm{t})}{\partial\boldsymbol{t}}\right) = 0 \tag{41}$$

Since the total charge (either quantum or classical) also satisfies a continuity equation (4), we conclude that the $free$ charge (due to electrons and the molecules understood as point charges) satisfies its own equation of motion

$$\frac{\partial}{\partial t}\langle\rho_{free}(\boldsymbol{x},t)\rangle + \nabla\cdot\langle j_{free}(\boldsymbol{x},t)\rangle = 0 \tag{42}$$

These results just show that the approximation developed in Sections 3.1 and 3.2 for the macroscopic charge and current densities are consistent among themselves. As expected, it confirms that our model of free particles does not create or destroy particles locally.

Such separation between $free$ and $notfree$ dynamics, cannot be translated into a separation between $free$ and $notfree$ displacement current. The divergence of Eq. (40) gives

$$\nabla\cdot\left(\langle\boldsymbol{j}_{free}(\boldsymbol{x},t)\rangle + \frac{\partial\mathbf{D}(\mathbf{x},\mathrm{t})}{\partial\boldsymbol{t}}\right) = \nabla\cdot\left(\langle\boldsymbol{j}_{free}(\boldsymbol{x},t)\rangle + \varepsilon_0\frac{\partial\mathbf{E}(\mathbf{x},\mathrm{t})}{\partial\boldsymbol{t}} + \frac{\partial\mathbf{P}(\mathbf{x},\mathrm{t})}{\partial\boldsymbol{t}}\right) = 0 \tag{43}$$

Therefore, in a two terminal device like the one in figure 1, we conclude that on some surfaces perpendicular to the transport direction, the total current is basically particle current, on other surfaces it is basically displacement current due to the time-dependent variations of the macroscopic $\mathbf{E}(\mathbf{x}, t)$, while on still other surfaces it is basically due to time dependent variations of the polarization $\mathbf{P}(\mathbf{x}, t)$, etc. On many surfaces, the current is just a mix of the three terms. In any case, this is the relevant message, the total current through any surface perpendicular to the transport direction of a two terminal device is equal.

This separation of particle current (flowing from one end of a device—say a resistor—to the other) and surface displacement current from the surface of the resistor conforms to time honored engineering practice. Physical resistors are typically represented as idealized Ohm's law resistances with an additional separate circuit element representing the sum of (1) the stray capacitance and (2) the displacement current on the (nonterminal) surfaces of the physical resistor[8]. Stray capacitors do not appear explicitly in descriptions of all electronic circuits (Horowitz and Hill 2015) but they are always implied and their practical importance is great, as is well explained on p. 581 of (Horowitz and Hill 2015). Successful devices depend on the proper control of stray capacitance (Johnson and Graham 2003, Scherz and Monk 2006).

'Stray capacitance' sounds as if it is a capacitance that could be removed if we were only clever enough to know how to do so. This is not the case, and no amount of work can reduce it beyond a minimum value. Stray capacitance is an unavoidable property of the electric field, describing the displacement current that is always present from the surface of real resistors. One might say stray capacitor holds the charge that is the "overhead", the price we must pay to create the potential across an ideal resistor. This overhead limits the speed in many practical devices, for example, it limits the refresh speed of the digital screens of our (large) televisions and computer terminals.

## 4. REALISTIC MACROSCOPIC DESCRIPTION OF THE CURRENTS

We move now to realistic descriptions of macroscopic systems. When Maxwell wrote his equations, technology did not allow measurement of time dependence at speeds faster than seconds and so delays between polarization and electric fields were essentially unknown. It was sensible then to begin study of the electric field by assuming that polarization was proportional to the electric field, with a single time independent constant embodied by a dielectric constant that is a real positive number, a constant. Polarization was supposed to be a local variable, independent of time or frequency, independent of the parameters and boundary conditions and even the positions of the boundaries and independent of the structure of the system.

**It is remarkable that the formulation of Maxwell that was developed entirely in a macroscopic context applies exactly also at the deep quantum level** (Albareda, Traversa et al. 2012, Marian, Zanghi et al. 2016) applied to atoms and within atoms to elementary particles, as shown in Section 2.3. One can only imagine what would

[8] A clear example is the ever popular metal film resistor, which is anything but a resistor at high frequencies due to its inherent inductive nature.

have happened if Maxwell had lived long enough to apply his electromagnetic field equations to the statistical mechanics he was helping to create (Garber, Brush et al. 1986).

Our technology today allows routine measurements in times less than $10^{-15}$ sec (Riek, Seletskiy et al. 2017), even in complex biological systems (Tsen and Tsen 2016), and our computations of atomic properties start at $2\times10^{-18}$ sec (Ferry, Goodnick et al. 2009, Vasileska, Goodnick et al. 2010), so it should not be a surprise that we resolve enormously more complex behavior of polarization charge than Maxwell. Indeed, it is safe to say that in the time scales just mentioned, polarization is never found to be characterized by a single dielectric constant (a single real positive number) in any material. And in most cases polarization depends on the parameters of the system, the boundary conditions, and their positions, and of course on the structure of the system. These are experimental facts known for nearly a century in many cases (Debye and Falkenhagen 1928, Debye 1929, Fröhlich 1958, Böttcher, van Belle et al. 1978, Buchner and Barthel 2001). It would seem wise then to use a formulation of Maxwell's equations that does not impose a fiction of a simple polarization property characterized by a dielectric constant that is a real positive number independent of time and frequency.

A hint of the complexities involved in real macroscopic systems can be found from the discussion of idealized harmonic oscillators given previously in this paper. Macroscopic systems involve myriads of interacting harmonic oscillators, and so obviously cannot be described by a simple polarization function. Serious attempts at derivation of polarization for simplified models of electron gases (Lundqvist and March 2013) show enormous complexity and applications to 'gases' made of quasi-particles in semiconductors p. 468-475 of (Mahan 1993) are hardly simpler.

Liquids have significantly more complex behavior than the idealized systems mentioned in the last paragraph. Liquids move in many more ways than solids, and movement is driven by multiple forces, diffusion and convection as well as temperature gradients, with diffusion being a crucial mechanism in most applications. Liquids are usually complex fluids and need to be analyzed by the mathematics of complex fluids, not ideal fluids or gases.

Ionic solutions and liquids are much more complex yet than 'uncharged' liquids—without permanent charge—because electric forces and migration in the electric field are dominant determinants of motion. Seawater resembles an ideal Ohm's law resistor much more than an uncharged liquid. Movements are driven by all fields in liquids and ionic solutions, everything is coupled to everything else, so ***polarization currents in these systems depend on all parameters and properties of all fields***, as well as on the structure and boundary conditions that constrain them.

In these systems, the distinction between bound charge and mobile charge is hard to make in a convincing way. Bound charge is found to have in phase components of current (in response to a sinusoidal perturbation over a range of frequencies) as well as the out of phase components characteristic of idealized bound charge and idealized polarization. Mobile charge is found to have out of phase components (in response to a sinusoidal perturbation over a range of frequencies) as well as the in phase components of idealized mobile charge of perfect conductors. Even the early simple models of polarization (Debye and Falkenhagen 1928, Debye 1929) have complex behavior. Polarization cannot be represented by a single dielectric constant, a real positive number independent of time or frequency in these oversimplified models. (See Historical Note early in this paper.) The (real positive) dielectric constant of the Maxwell equations becomes a complex variable (with real and imaginary parts, magnitude and phase) in the Debye model of polarization. As these models are adapted to deal with real systems, the approximation of polarization by a single dielectric constant becomes worse and worse.

Looking at real systems from the point of view of the experimental scientist—who does not know ahead of time what mechanism produces out of phase or in phase components of currents—it seems a daunting task to determine whether an in-phase component of current arises from a lag in a nonideal polarization current produced by complex movements of bound charge, or from a conduction current. It is difficult and, in our opinion, obviously artificial to make a distinction from experimental data alone, between nonideal properties of polarization current (of bound charges) and nonideal properties of conduction currents (of mobile charges).

For these reasons we follow the lead of (Purcell and Morin 2013, section 10.4, p. 505-507) and abandon the isolation of polarization current, but rather deal with any type of current at all, isolating only the vacuum displacement current (see eq. 31.1) that can in fact be characterized exactly by a single real constant the permittivity of free space $\varepsilon_0$. We write current in any material as it is written for a vacuum in most textbooks of electrodynamics. We return to more traditional descriptions later to maintain contact with the traditional literature.

We find that abandoning the traditional approach is disturbing to our colleagues, so we think it necessary to cite others who have this view. In the well-known textbook Purcell and Morin p. 507 of (Purcell and Morin 2013) write

> "…. in the real atomic world the distinction between bound charge and free charge is more or less arbitrary, and so, therefore, is the concept of polarization density **P**. The molecular dipole is a well-defined notion only where molecules as such are identifiable – where there is some physical reason for saying, 'This atom belongs to this molecule and not to that.' In many substances such an assignment is meaningless. An atom or ion may interact about equally strongly with all its neighbors; one can only speak of the whole…."

A liquid, or an ionic solution, fits perfectly into Purcell and Morin's discussion. The structure of liquids (see Section 23.2 p. 629 of the definitive text (Berry, Rice et al. 2000)) ensures that "we cannot isolate any one pair of molecules from interactions with other molecules" (p. 529). Everything interacts with everything else. Analysis in terms of a single distribution function $W(\boldsymbol{x})$ is not likely to be adequate in a system like that, a liquid or an ionic solution.

Quotations aside, the reason to abandon the traditional approach is clear simply from the properties of the distribution function used in classical analysis. The distribution function $W(\boldsymbol{x})$ is written with one functional dependence, only on $\boldsymbol{x}$. It should be immediately obvious that a single function $W(\boldsymbol{x})$ with functional dependence only on $\boldsymbol{x}$ is unable to deal with the enormous range of dielectric properties observed experimentally in equilibrium measurements of linear dielectrics, for nearly a century, (Debye and Falkenhagen 1928, Debye 1929, Onsager 1936, Oncley, Ferry et al. 1940, Oncley 1942, Fuoss 1955, Fröhlich 1958, Van Beek 1967, Nee and Zwanzig 1970, Hubbard, Onsager et al. 1977, Böttcher, van Belle et al. 1978, Anderson 1994, Barthel, Buchner et al. 1995,

Barthel, Krienke et al. 1998a, Buchner and Barthel 2001, Pitera, Falta et al. 2001, Oncley 2003, Prodromakis and Papavassiliou 2009). These measurements are now called impedance or dielectric spectroscopy (Macdonald 1992, Kremer and Schönhals 2003, Barsoukov and Macdonald 2005). Their main topic is the complex functional dependence of dielectric behavior that cannot be described by a single dielectric constant, a real positive number.

Non-equilibrium systems have much richer behavior than the equilibrium systems studied in impedance or dielectric spectroscopy. Indeed, that is exactly why most of the devices and machines of our technology are non-equilibrium, as are all of the systems of life. The polarization of non-equilibrium systems can also not be described by theories involving a single distribution function $W(\boldsymbol{x})$ with functional dependence only on $\boldsymbol{x}$. Our technology and much of biology involve devices with well-defined inputs and outputs, as well as robust input output relations. Devices obviously include variables and parameters to describe inputs and outputs. These variables describe the essential function of devices. If the variables are not present in a description of polarization at all, the description obviously cannot describe how polarization changes as the inputs and outputs change. $W(\boldsymbol{x})$does not contain variables to describe inputs, outputs.

It might seem at this juncture that the situation is desperate and nothing useful can be said about systems in general, because the properties of polarization are so diverse, and that would certainly be the appropriate conclusion if only mechanical and steric forces were involved.

The remarkable result is that something can be said, and what can be said is very powerful indeed, because of the special properties of the electric field, because of Maxwell's displacement current, that occurs in electrical problems in a special way.

***Conservation of current and thus Kirchoff's current law does not depend on any discussion of polarization.*** It is true at the fundamental quantum level as shown in Section 2.3 and it is true everywhere else as well.

Kirchoff's current law is (nearly) enough to analyze and synthesize the linear and nonlinear networks of electronic devices, passive and active because those circuits have simple structure. They are fundamentally one dimensional systems with branches. Kirchoff's current law is (nearly) enough to analyze and synthesize our electronic technology, digital and analog, that has allowed a $10^9$ improvement in functionality in 60 years.

## 4.1 Mathematics Of Current Flow.

A crucial property of the electric field can be derived without mention of polarization at the quantum level as we have shown already and in general (Mansuripur and Zakharian 2009, Eisenberg 2016a, Eisenberg 2016b) as we shall see. Conservation of total current $\mathbf{J}_{total}$ and thus Kirchoff's law for total current (in one dimensional branched systems) can be derived without mention of polarization. The mathematical derivation is quite succinct, although the physical meaning of that derivation seems to produce lengthy discussion.(Eisenberg 2016c)

The mathematical derivation depends on one of the key equations of electrodynamics, Ampere's law, as modified by Maxwell.[9] For easier reading, we rewrite equations (1) and (6.4) again here. We use capital letters, but we understand them without the spatial average discussed in section 3.1. They are fundamental and universal laws true on all scales, within and between atoms and true on macroscopic scales as well.

$$\frac{1}{\mu_0}\nabla\times\boldsymbol{B} = \mathbf{J}_{total} = \mathbf{J}_D + \mathbf{J}_Q; \qquad \mathbf{J}_D = \varepsilon_0 \frac{\partial \mathbf{E}}{\partial t} \tag{43}$$

As already mentioned, $\mathbf{J}_Q$ describes all movements of charge associated with matter, in this formulation of Ampere's law (see p. 276 of and Ch.3. of (Lorrain and Corson 1970)). $\mathbf{J}_D$ describes properties of the vacuum—i.e., free space—and is independent of the properties of matter. Polarization properties of matter are included in $\mathbf{J}_Q$ as advocated in the quotation cited above from p. 507 of (Purcell and Morin 2013). The historical discussion of (Arthur 2013) makes it easier to abandon traditional representations of polarization and **D** fields because it makes clear that they were never based on experimental reality. Eisenberg (2016a, 2016b) uses traditional representations of polarization to connect this approach to the traditional literature on linear dielectrics, used to describe the complex behaviors of polarization and $\mathbf{J}_Q$ found in experiments.

**Universal Law.** It is now a simple step to a universal law for current flow true for any polarization property at all. We apply the vector identity $\nabla\cdot(\nabla\times\boldsymbol{B}/\mu_0) = 0$ and derive conservation of total current using a realistic description of macroscopic materials, as we did for atomic scale particles in expressions (7) and (8) in section 2.2.

$$\nabla\cdot\left(\mathbf{J}_{total}\right) = \mathbf{0} \qquad \nabla\cdot\left(\mathbf{J}_Q + \overbrace{\varepsilon_0\frac{\partial \mathbf{E}}{\partial t}}^{\mathbf{J}_D}\right) = 0 \tag{44}$$

Conservation of total current $\mathbf{J}_{\boldsymbol{total}}$ is possible because the electric field **E** field changes according to Ampere's law. The key physical idea is that the **E** field is a variable that changes the displacement current $\mathbf{J}_D$ so $\mathbf{J}_{\boldsymbol{total}}$ is conserved.

Conservation of $\mathbf{J}_{total}$ is **universal**, derivable for particles on the atomic scale (see Section 2.2) or for macroscopic systems without mention of the polarization or dielectric properties of matter.

We write a simple approximation derived from eq. (44) that shows one way the electric field **E** can change its shape—i.e., how it depends on time—to ensure conservation of current.

[9] Historically, this equation was a fulcrum in the history of physics: it allows waves to propagate at a velocity $c$ (units: meter/sec) determined entirely by constants describing the strength of the magnetic field $\mu_0$ (units: henry/meter) and the electric field $\varepsilon_0$ (permittivity of free space, farads/cm), namely $c = 1/(\mu_0\varepsilon_0)^{\frac{1}{2}}$. Measurements of electrical and magnetic phenomena are enough to correctly calculate the speed of light!

If the electric field changes according to the equation

$$\mathbf{E} = -\int_{o}^{t}\left(\mathbf{J}_Q(t';_{\,etc.})/\varepsilon_0\right)dt', \tag{45}$$

current is conserved. Eq. (45) is obviously not a general statement. Eq. (45) implies eq. (44) but eq. (44) does not imply eq.(45). An explicit general statement for how **E** must change to satisfy Ampere's law and Maxwell's equations is much more complicated

## 4.2 Conservation of Charge.

Conservation of current is closely connected to conservation of charge (see discussion in section 2.1), through the continuity equation, which we now derive using the Gauss equation of electrostatics, often called Maxwell's first equation in (6.1) rewritten here as:

$$\boldsymbol{\nabla}\cdot\ \mathbf{E} = \frac{\rho_Q}{\varepsilon_0} \tag{46}$$

Here $\rho_Q$ describes the density of all charge associated with the density of mass $\rho_{mass}$. The charge density $\rho_Q$ includes *(i)* any charge distribution independent of the electric field, *(ii)* polarization charge of perfect dielectrics (characterized by a single dielectric constant that is a real positive unchanging number), and *(iii)* any other charge that depends on the electric field, whether the dependence is simple as in the polarization charge, or more complicated, depending (for example) on other fields. The dependence of charge on other fields is the key to understanding many phenomena in complex fluids (Doi and Edwards 1988, Hou, Liu et al. 2009, Liu 2009, Hyon, Kwak et al. 2010); electrorheology (Sheng, Zhang et al. 2008, Zhang, Gong et al. 2008), for example, of the Marangoni effect (Velarde 2003, Hu and Larson 2005, Sun, Liu et al. 2009), and 'tears of wine' (Fournier and Cazabat 1992) and 'oil on water' , studied by B. Franklin (Franklin, Brownrigg et al. 1774); electrodiffusion models like the (Poisson) drift diffusion equations (Van Roosbroeck 1950, Gummel 1964, Macdonald and Franceschetti 1978, Selberherr 1984, Markowich, Ringhofer et al. 1990, Jerome 1995) called Poisson Nernst Planck (PNP) equations in electrochemistry and biophysics (Eisenberg and Chen 1993, Eisenberg 1996, Eisenberg 1999, Coalson and Kurnikova 2005, Ji, Liu et al. 2015).

Now, we differentiate Gauss' equation (46) with respect to time, and interchange order of differentiation in time and space, on the way to deriving the continuity equation for charge density $\rho_Q$

$$\boldsymbol{\nabla}\cdot\left(\varepsilon_0\frac{\partial}{\partial t}\mathbf{E}\right) = \frac{\partial\rho_Q}{\partial t} \tag{47}$$

but from eq. (4)

$$\boldsymbol{\nabla}\cdot\left(\varepsilon_0\frac{\partial\mathbf{E}}{\partial t}\right) = \overbrace{\boldsymbol{\nabla}\cdot\ (\boldsymbol{\nabla}\times\boldsymbol{B}/\mu_0)}^{=\mathbf{0}}\ -\boldsymbol{\nabla}\cdot\ \mathbf{J}_Q \tag{48}$$

so we have the continuity equation relating the flux of any mass carrying charge to the density of that mass.

$$\nabla \cdot \mathbf{J}_Q = -\frac{\partial}{\partial t}\rho_Q \tag{49}$$

Note the electrical field $\mathbf{E}$ and the displacement current $\mathbf{J}_D = \varepsilon_0\, \partial\mathbf{E}/\partial t$ do not enter into the continuity equation. Both the flux $\mathbf{J}_Q$ and the charge density $\rho_Q$ describe all charge, whatever its origin.

We now describe some of the many forms of charge, hoping to connect the reader to the more classical literature in this way and to motivate the reader to abandon the use of a dielectric fiction, namely a dielectric constant that is a single real number independent of time, frequency, and all other variables and fields.

**(1) Perfect idealized dielectrics** $\boldsymbol{J}_Q$ of a perfect dielectric includes polarization charge that is well described by a dielectric constant that is a positive real number that never varies with anything. Perfect dielectrics possess the idealized polarization charge of classical textbooks, reaching back to 1893, as described in (Becker and Sauter 1964), see (Abraham and Becker 1932). The idealization is an important aid in teaching and exploratory analysis of new systems, because it allows simplified theories.

Perfect dielectrics have (1) zero current flow as $t \to \infty$ when steady voltage is applied and (2) 90 degree phase difference between current and voltage at ***all*** frequencies when sinusoids are studied (3) amplitude (and phase) of current/voltage independent of frequency when sinusoidal voltage/current is applied.

**(2) Perfect idealized conductors** have zero phase difference between current and voltage at all frequencies when sinusoids are studied. Current and voltage are proportional to each other, with a proportionality constant that is a single real positive constant at all times.

It should be clearly understood, however, that matter never behaves as a perfect dielectric, with idealized polarization, or perfect conductor over the range of times and conditions of technological, biological, or chemical interest, as documented at length previously in this paper. Real materials are neither dielectrics nor conductors but rather a combination of both, with properties that always vary dramatically with time, and often with many other variables.

**(3) Linear Dielectrics** are linear in the electric field, meaning currents are strictly proportional to the strength of the electric field at each time and position. The electrical potential (or current) can then be "divided out" and the linear dielectric can be characterized by properties and parameters that do not depend on voltage or current, parameters like conductance, resistance, capacitance, dielectric coefficient, admittance, impedance, and reactance. Linear dielectrics have properties that vary dramatically with frequency/time, composition, and concentration of the chemical species that make up the dielectric as shown in measurements done for nearly a century in a huge literature now called impedance spectroscopy (Debye and Falkenhagen 1928, Debye 1929, Onsager 1936, Oncley, Ferry et al. 1940, Fuoss 1955, Fröhlich 1958, Van Beek 1967, Nee and Zwanzig 1970, Hubbard, Onsager et al. 1977, Böttcher, van Belle et al. 1978, Anderson 1994, Barthel, Buchner et al. 1995, Barthel, Krienke et al. 1998a, Buchner and Barthel 2001, Pitera, Falta et al. 2001, Barsoukov and Macdonald 2005, Prodromakis and Papavassiliou 2009). The literature includes many special effects (Debye Falkenhagen; Maxwell Wagner, for example) that highlight the complexity of phenomena. Every linear dielectric has properties that change dramatically with time or frequency, without exceptions known to us.

**(4) Materials in general.** In most materials and all ionic solutions, $\boldsymbol{J}_Q$ includes coupled, often ***nonlinear*** properties that cannot be comfortably described by classical theory but seem to require a more general description. In fact, the coupled properties of ionic solutions have not yet been successfully described (Zemaitis, Clark et al. 1986, Barthel, Buchner et al. 1995, Barthel, Krienke et al. 1998a, Jacobsen, Penoncello et al. 2000, Myers, Sandler et al. 2002, Wilczek-Vera and Vera 2003, Lin, Thomen et al. 2007, Tresset 2008, Kontogeorgis and Folas 2009, Fraenkel 2010, Hünenberger and Reif 2011, Eisenberg 2013b, Liu and Eisenberg 2015, Rowland, Königsberger et al. 2015, Kohns, Reiser et al. 2016, Wilczek-Vera and Vera 2016, Xie, Liu et al. 2016) over a range of compositions and concentrations found in seawater and living organisms (Kunz 2009, Kunz and Neueder 2009) even at equilibrium (without flows of any kind).

Nonlinear properties characterize most transport in biology (Cole 1972, Ruch and Patton 1973a, Ruch and Patton 1973b, Weiss 1996, Keener and Sneyd 1998, Ashcroft 1999, Hille 2001, Jackson 2006, Boron and Boulpaep 2008, Koeppen and Stanton 2009, Prosser, Curtis et al. 2009, Gabbiani and Cox 2010, Zheng and Trudeau 2015) and cannot easily be described by generalizations of the permittivity $(\varepsilon_r - 1)\varepsilon_0$ despite the attempts of Cole (Cole and Curtis 1936, Cole and Curtis 1938, Cole and Curtis 1939, Cole 1947, Cole 1972, Huxley 1992). Currents in macroscopic biological systems (Hodgkin and Huxley 1952a, Hodgkin and Huxley 1952b, Hodgkin and Huxley 1952c, Huxley 2000, Huxley 2002) and in the molecules producing and controlling the currents (Armstrong and Bezanilla 1973, Bezanilla, Vergara et al. 1982, Bezanilla 1985, Vandenberg and Bezanilla 1991, Sakmann and Neher 1995, Neher 1997, Bezanilla and Stefani 1998, Vargas, Yarov-Yarovoy et al. 2012, Horng, Eisenberg et al. 2017) are described by nonlinear differential operators including terms quite different from $(\varepsilon_r - 1)\varepsilon_0\,\partial\mathbf{E}/\partial t$, called the Hodgkin Huxley equations when the currents are macroscopic (op. cit.). Quite different representations are needed for currents that flow through single protein channels (Sakmann and Neher 1995, Neher 1997).

Nonlinear charge movements—some extremely nonlinear (Wegener 2005)—create nonlinear optics, studied initially as lasers (Sutherland 2003, Boyd 2008, Hill and Lee 2008). Extraordinary optical devices are possible if materials are built with spatial variations of displacement current on the atomic scale, creating the exciting areas of photonics, quantum chiral optics (Lodahl, Mahmoodian et al. 2017) and cloaking devices (Islam, Faruque et al. 2016, Zheng, Madni et al. 2016).

Spatially dependent nonlinear charge movements are creating several of the new fields of science and technology we read about in newspapers. Basov and Folger (Basov and Fogler 2017) write “High-temperature superconductivity, unconventional magnetism, and charge-ordered states are examples of the spectacular properties that arise in solids through many-body effects, a consequence of electrons strongly interacting with one another and with the crystal lattice” Lundeberg et al, point to the future (Lundeberg, Gao et al. 2017) “The response of electron systems to electrodynamic fields that change rapidly in space is endowed by unique features, including an exquisite spatial nonlocality.” Dielectric fictions are left far behind in this work.

### 4.3 Flow of mass.

The understanding of dynamics of charge movement $\mathbf{J}_Q$ depends of course on the dynamics of mass $\mathbf{J}_{\boldsymbol{mass}}$ . A usable model requires explicit connection between the equations of motion of mass and charge, as for example, in the charged harmonic oscillators discussed earlier (Hall and Heck 2011). We consider a number of systems to get a feel for the issues involved.

Consider first the flow of uncharged matter, the traditional subject of fluid mechanics, and theory of complex fluids. If the mass has no charge (of any kind under any conditions), its flow is specified by a mixture of conservation of mass and constitutive equations. In simple cases, field equations as complex as the Navier Stokes equations arise. But mass is moved by many forces, for example, pressure, and temperature gradients depending on frictional dissipative processes. $\mathbf{J}_{mass}$ involves multifaceted interactions of various fields and differential equations, just as does $\mathbf{J}_Q$. Each facet of the various fields can interact with every other. Fitting parameters appear in the numerous cross terms of the differential equations describing these interactions and these are often determined poorly by experimental work.

A variational approach minimizes the number of fitting parameters and leads to transferrable models useful in the design of devices. The variational treatment guarantees that results are mathematically consistent, with all variables satisfying all field equations and boundary conditions, with a minimal set of fitting parameters, that are in fact constant when the model fits data successfully. The ***EnVarA*** formulation introduced by Chun Liu, more than anyone else, is such an approach, including dissipation, as it must when condensed phases are involved (Ryham, Liu et al. 2006, Ryham 2006, Eisenberg, Hyon et al. 2010, Horng, Lin et al. 2012, Forster 2013, Wu, Lin et al. 2014b, Wu, Lin et al. 2014a, Xu, Sheng et al. 2014, Wu, Lin et al. 2015, Wang, Liu et al. 2016). Movements in any condensed phase involve strong atomic interactions on the $10^{-17}$sec time scale (‘collisions‘) because condensed phases have little empty space, by their very definition. Friction and dissipation are the macroscopic results of collisions. Treatments of condensed phases, including liquids and ionic solutions must include friction if they are to deal with flow.

## 4.4 Flow of uniformly charged matter.

This simple kind of matter has a constant density of charge (per density of matter). The charge density is permanent, independent of the local electric field, and distributed uniformly in space. The description of uniformly charged matter requires variational methods just as does the flow of uncharged matter.

It is unusual—if not unheard of—for the charge density of matter to be constant independent of the local electric field as we assume here. The electric field is so strong, as we have discussed, that it nearly always distorts matter, creating positive and negative poles of charge, leading to the name polarization for the change in the spatial distribution of charge induced by the electric field.

Matter usually consists of molecules that have themselves asymmetrical permanent distribution of charge produced by a combination of polar bonds and asymmetrical distribution of permanent charges like the acid and base groups of amino acids, or other weak acids or bases. Asymmetrical polar molecules like these rotate in electric fields including the fluctuating fields produced by thermal motion of charged atoms and molecules. Polar molecules have complex Brownian motion, involving rotation and translation, so the averaged distribution of charge depends on frequency or time, temperature, and the electric field itself, as well as of course any permanent charges, or ions with permanent charge that are present, as they usually are. More general molecules have stretching motions as well as complex twisting motions, not easily described in a general way, certainly not as elasticity. A brief look at the structure of nucleic acids and how they wind, unwind, as they self-assemble into ribosomes or chromosomes shows how complex these motions can be. (Remember that DNA and RNA are characterized by very large densities of acid groups, with their permanent negative charge on carboxylates, as well as by the strongly polar bonds of their nucleobases, purines and pyrimidines with large permanent partial charges, e.g., nearly $-0.3e$ on the oxygen of carbonyls.)

The flow of charged matter in general is thus very complex indeed. Charged molecules are polarized by the electric field as just described. The charges of the molecules also help create the electric field of course. Everything interacts with everything else and all relevant equations must be solved together. Consistently, with all variables satisfying all equations under all conditions, with one set of unchanging (and thus transferable) parameters.

The flow of mass $\mathbf{J}_{mass}$ and the flow of charge $\mathbf{J}_Q$ depend individually on the electric field in an intricate way, as we have discussed. The variable that relates these flows is the charge per mass, and that too has complex properties, as charged molecules, stretch, rotate, and interact. 'Everything depends on everything else' in these systems. Variational methods keep track of these interactions, in our view, and are particularly useful because they guarantee that all the output (dependent) variables satisfy all equations and boundary conditions.

The flow of $\mathbf{J}_Q$ is more complex than the flow of uncharged matter because the electric field strongly interacts with all the fields and flows of the variational treatment. The electric field is remarkably strong and so the electrical terms are large—often dominant—even in systems that are uncharged on the average.

Consider an uncharged system like liquid argon (Hirschfelder, Curtiss et al. 1964). The fluctuations in charge density in systems with zero mean charge like liquid argon produce dispersion forces (Israelachvili 1992, Parsegian 2006, Stone 2013)) that dominate the properties of the liquid argon and are of important components of all intermolecular forces.

Consider the technologically important phenomenon of dielectrophoresis (Pohl 1978, Jones 2003) used in the separation of chemically similar molecules. In dielectrophoresis, particles with zero permanent charge can be transported by the electric field because the particles have induced polarization charge. That is to say, in formal terms, $\frac{\partial^2 \mathbf{E}}{\partial x^2} \neq 0 \Rightarrow$ flow by dielectrophoresis. Phenomena like dielectrophoresis produce ***both*** transport of $\mathbf{J}_{mass}$ and $\mathbf{J}_{\mathrm{Q}}$ even when the molecules involved have no net charge.

Each of these systems requires a separate model and entire professions are devoted to each type of model. Few universals exist, but where they exist they are most helpful in constructing and constraining models. Conservation of charge, conservation of mass are such universals. We believe conservation of current $\mathbf{J}_{\mathrm{total}}$ is another universal that will be helpful in constructing models.

Conservation of current has rarely been used as an independent constraint on models probably because the current conserved is usually taken as the flux of charge $\mathbf{J}_{\mathrm{Q}}$ that depends on the dielectric properties of matter. See however (Mansuripur and Zakharian 2009) and other extensive discussions of displacement current (Zapolsky 1987, Arthur 2008, Selvan 2009, Arthur 2013). Dielectric properties, and polarization in general, are drastically oversimplified in usual treatments. Laws of current flow that involve these oversimplifications are distrusted, for good reason, and so investigators do not use those laws when they try to construct realistic models of real matter.

We hope we have convinced the reader that conservation of electrical current $\mathbf{J}_{\mathrm{total}}$ is an independent constraint just as much as conservation of charge $\rho_Q$ and conservation of mass. $\mathbf{J}_{\mathrm{total}}$ is conserved because it includes Maxwell's displacement current. That current is not included in the usual descriptions of mass and its flow and so conservation of current $\mathbf{J}_{\mathrm{total}}$ cannot be derived from the conservation laws of mass and its flow.

Conservation of current arises because of the special properties of the electric field and its displacement currents. Ampere's law eq. (4) guarantees that conservation of mass $\rho_{mass}$ and its flow $\mathbf{J}_{mass}$ does not imply conservation of total current. We believe conservation of total current is a universal property of the electric field, from atoms to animals, that does not involve polarization or its properties.

### 4.5 Conservation of current in electronic technology.

In the branched one dimensional circuits of our electronic technology, conservation of $\mathbf{J}_{total}$ implies (Bhat and Osting 2011) Kirchoff's 'current' law, where 'current' is $\mathbf{J}_{total}$ not $\mathbf{J}_Q$. All the $\mathbf{J}_{total}$ that flows into a node flows out, as described by Kirchoff's current law. $\mathbf{J}_{total}$ **is never stored, not even a little bit, not at any time, not at any place.**

In contrast to the flow of current, the flow of charge is ***not*** described by Kirchoff's law. All of the current $\mathbf{J}_Q$ that flows into a node does ***not*** flow out. According to eq.(45), some of the current $\mathbf{J}_Q$ is stored to create $\mathbf{E} = -\int_0^t \left(\mathbf{J}_Q(t'; etc.)/\varepsilon_0\right) dt'$ and that $\mathbf{E}$ is exactly what is needed to enforce Kirchoff's 'current' law, where 'current' is $\mathbf{J}_{total}$, ***not*** $\mathbf{J}_Q$.

The stored charge taken from $\mathbf{J}_Q$ can be said to be 'stored in the capacitance of free space' determined by $\varepsilon_0$ and the geometry of the system. The stored charge taken from $\mathbf{J}_Q$ does not appear explicitly in most descriptions of electronic circuits (Horowitz and Hill 2015) because it is often viewed as a 'parasitic' stray capacitance, something to be avoided and denied, like other stray parasites. But every engineer knows that parasitic capacitance is important in the practical implementations of circuits p. 581 of (Horowitz and Hill 2015) and successful devices depend on the proper control of stray capacitance (Johnson and Graham 2003, Scherz and Monk 2006).

Stray capacitance is clearly an unavoidable property of the electric field equation (1) that can produce $\mathbf{E} = -\int_0^t \left(\mathbf{J}_Q(t'; etc.)/\varepsilon_0\right) dt'$ by storing charge. That stored charge and that $\mathbf{E}$ is exactly what is needed to enforce Kirchoff's 'current' law, where 'current' is $\mathbf{J}_{total}$. 'Current' is not $\mathbf{J}_Q$.

As we have carefully stated earlier, leaving the stray capacitance out of idealized circuits is a well-motivated over-simplification making it easier to teach circuit theory to newcomers who have not actually built circuits. But that simplification produces inconsistencies if Kirchoff's current law is mistakenly applied to the current $\mathbf{J}_Q$. Kirchoff's law for $\boldsymbol{J}_{total}$ is consistent with Maxwell's equations. Kirchoff's law for $\mathbf{J}_Q$ is not consistent with Maxwell's equations, if circuits omit the stray parasitic capacitance of free space that supports displacement current $\varepsilon_0\, \partial\mathbf{E}/\partial t$.

The conservation of current is most striking in a series circuit. In a series circuit, $\mathbf{J}_{total}$ is equal everywhere, no matter what the physics of current flow in each component (see examples in Fig. 2 of (Eisenberg 2016c)). Note the currents $\mathbf{J}_{total}$ are equal at any time, including at the atomic scale $10^{-16}$ sec. Currents of $\mathbf{J}_Q$ are certainly not equal on the atomic scale because field fluctuations $\partial\mathbf{E}/\partial t$ are so large on the atomic scale, producing huge displacement currents $\mathbf{J}_D = \varepsilon_0\, \partial\mathbf{E}/\partial t$ in any consistent simulation of atomic or molecular dynamics. See the general review of computational electronics (Vasileska, Goodnick et al. 2010).

### 4.6 Conservation of current in chemistry.

Chemical reactions are described as a series of reactions that obey the law of mass action. Reactions involving charged reactants produce current flow. It was a surprise (Eisenberg 2014a,b) to find that models of series chemical reactions $A \rightarrow B \rightarrow C$ have unequal currents $I_{AB} \neq I_{BC}$. The current $A \rightarrow B$ is not constrained to equal the current $B \rightarrow C$ in classical chemical models. The models are usually not transferable. The descriptions of chemical reactions typically require different rate constants under different experimental conditions and so have limited utility. In future work, we will try to modify the description of chemical reactions so they conserve current.

Chemical reactions involve charge storage as well as the flux of charge. Maxwell's equations, and their displacement current, are needed to describe that storage of charge, as we have seen. In the chemical literature, stored charge is often described by the Born equation (Atkins and MacDermott 1982) for self-energy in an idealized systems without boundary conditions. For example, the interactions of ions with water ('solvation') are widely described by the Born equation, particularly in proteins and macromolecular systems (Bashford and Case 2000). The Born equation does not allow current flow, does not deal with displacement current in general, and ignores the boundary conditions that can change the qualitative features of the electric field in practically important ways (Mertens and Weeks 2016). The Born equation is a drastic approximation to the complexities of current flow in chemical reactions and systems.

Higher resolution analysis involving simulations on the atomic scale are performed widely in molecular biology because of the wonderful structures (of more than $10^5$ proteins, typically made of $>10^5$ of atoms) available mostly from x-ray crystallography. The beauty and power of these structures has enormous appeal to the mind's eye, but that appeal makes it easy to overlook the other demands of the mind.

Protein structures do not include the electrical potentials and macroscopic concentrations that power the currents that flow throughout living systems, and therefore simulations are needed. Protein structure has allowed us to identify and look at the atoms that make up the proteins of life but structures are not enough. One can learn a great deal from snapshots of an automobile engine and its pistons. But one needs to study the motions to know how the engine works.

Atomic resolution simulations extend our knowledge of protein structures in most important ways. But they do not provide an easy extension from the atomic time scale $2\times10^{-16}$ sec to the biological time scale of gating currents that starts at $50\times10^{-6}$ sec and reaches $5\times10^{0}$ and longer (we hope). Calculations of currents from simulations must average the trajectories of atoms that last $50\times10^{-3}$ sec and are sampled every $2\times10^{-16}$ sec) involving some $10^6$ atoms all of which interact through the electric field to conserve charge and current, while conserving mass. Simulations like molecular dynamics do not provide an easy treatment of interactions. It is obviously impossible to simulate all the interactions of the tremendous number of particles involved and their interactions which are so numerous that the word 'tremendous' seems quite inappropriate. (Some $10^{21}$ atoms are involved and interactions are not just pairwise, because of the crucial role of polarization. Polarization ensures that forces between any pair of atoms depend on the locations of all other atoms. Thus, the total number of interactions is far larger than $10^{21}$factorial!)

It is difficult to enforce continuity of current flow in simulations of atomic dynamics because simulations compute only local behavior while continuity of current is global, involving current flow far from the atoms that control the local behavior. It is impossible to enforce continuity of current flow in calculations that assume equilibrium (zero flow) under all conditions. Current cannot be both zero and finite. Periodic boundary conditions are widely used in simulations. Such conditions take a box of material and replicate it identically, so the potential at the corresponding edges of the box are identical. If the potentials are identical, current will not flow. Periodic boundary conditions of this sort are incompatible with current flow from one boundary to the other. Voltage clamp experiments, and natural biological function involve current flow from one boundary to another. Atomic resolution simulations of current flow are not feasible now nor is it likely they will ever be feasible when trace ions (like $Ca^{++}$) are involved, as they are in most biological systems. Too many water molecules must be computed to determine the trace concentration of $Ca^{++}$.

It seems to us that the wonderful resolution of structure and atomic simulation must be combined in a hierarchy of models so we can understand how changes in a handful of atoms control macroscopic current flow in proteins and biology. Continuum models are needed to extend high resolution simulations to macroscopic reality.

Continuum models compute current flow as it depends on a variety of conditions, namely different electrical potentials, different concentrations and compositions of ionic solutions, and different structures of confining systems. The quantities from computations/analyses of models can be compared directly with experimental measurements of current. The quantitative models are dramatically reduced in complexity compared to structures or simulations of structures in atomic detail, but they are precise. Such is the nature of most physical models of condensed phases. Such must be the nature of physical models of biological function, in our view.

## 5. Conclusion

### Atomic Control And Displacement Current

A few atoms control the transistors of our computers. A few atoms control living systems, although these atoms are billions of times smaller, and move thousands of millions of times faster than living things. Somehow the atoms do manage macroscopic control. How is this possible?

We need experiments, models, mathematics and simulations to approach an answer to this question. No single approach will succeed itself, despite the near-sighted vision of scientists who know and seek to support only their own approach. A nested hierarchy of models, at different length and time scales, are needed to connect the atoms to the macroscopic world of life and computer chips. Mathematics and simulations are needed to compute what these models can do and compare the computations with experiments.

Implementing these ideas in our models is hard to do. Reaching to the macroscopic scale, we develop models with lower resolution, and coarser grain, as presented in Sec. 3. But it is easy to lose significant fine structure of the atomic scale by the very process of coarse graining. Some atomic details matter a great deal, but most atomic details do not matter at all.

It is perhaps possible to construct the hierarchy of nested models one step at a time with exhaustive experimentation accompanied by theory and simulation at every stage. Indeed, that is the approach used (for the most part) in constructing the nested hieerarchy of transistors, integrated circuits, logic, arithmetic, and memory management units that make our computers.

But much of science is analysis, not design. Much of science, and most of biology, is concerned with the inverse problem of determining how something works, from measurements of inputs and outputs, using independent knowledge of power supplies and structure. Such backwards engineering is made much easier if there are principles and laws that apply widely in systems of diverse structure and scale.

The laws of electricity are true on all scales. The great majority of our technology, and all our information technology, depend on these laws and their ability to transfer understanding developed on one scale to other scales. The laws of electricity are true on all scales with one set of parameters that do not change. We imagine that the universal nature of these laws allows atoms to control the macroscopic functions of life and computers, although we are quite aware of the gap between our imagination and proven truth.

Conservation of current is a law we focus on here because that conservation law extends throughout space and couples ‘everything to everything else’ in a more dramatic way than other conservation laws. It is true on the atomic scale, within atoms, and between stars. On the macroscopic scale of life, conservation of current necessarily links far separated boundaries to each other, connecting inputs and outputs to one another, and thereby creating devices.

We show that conservation of current is exact in systems with such complex charge movements that the words dielectric and polarization are not useful. Displacement current remains defined precisely and exactly even in such systems. Maxwell’s displacement current allows conservation of current to be true universally from atoms to stars. We suspect that Maxwell’s displacement current flows from atomic to macroscopic scales and helps evolution find groups of atoms that can control machines and organisms, although our suspicion is certainly not settled science. Our suspicion is a guess, a reach, far beyond our grasp.

We believe models, simulations, and computations should conserve current on all scales, as accurately as possible, because physics conserves current that way. We believe models will be much more successful if they conserve current at every level of resolution, the way physics does. We surely need successful models as we try to control macroscopic functions by atomic interventions, in technology, life, and medicine.

Maxwell’s displacement current lets us see stars. We hope it will help us see how atoms control life.

## Acknowledgment

Xavier Oriols thanks the Fondo Europeo de Desarrollo Regional (FEDER) and the 'Minister de Cynical e Innovación' through the Spanish Project TEC2015-67462-C2-1-R, the Generalists de Catalunya (2014 SGR-384), the European Union's Horizon 2020 research and innovation program under grant agreement No 696656.

Bob Eisenberg thanks Stuart Rice for his important role motivating this work. During a seminar of Bob's in May 1997, in the Chemistry Department at the University of Chicago, Stuart pointed out that 'electric field equations apply at all scales', provoking thoughts that led to (parts of) this paper, many years later.